\documentclass[useAMS,usenatbib]{mn2e}
\usepackage{graphicx}
\usepackage{amsmath}
\usepackage{amssymb}
\usepackage{amsfonts}
\usepackage{natbib}
\usepackage{graphicx}
\usepackage{epsfig}

\usepackage[caption=false]{subfig}
\usepackage{url}
\usepackage{textcomp}
\usepackage[squaren]{SIunits}
\usepackage{float}
\usepackage[T1]{fontenc} 
\usepackage{aecompl}
\def\fig{Figure}

\def\Fig{Figure}
\def\Figs{Figures}
\def\sect{Sect.}

\def\Sect{Section}

\def\tab{Table}

\def\Tab{Table}
\def\Tabs{Tables}
\def\eqn{Equation}

\def\Eqn{Equation}

\def\deg{$^{o}\,$}
\def\arcm{$^{\prime}\,$}
\def\arcs{$^{\prime\prime}\,$}
\def\mujybm   {${\rm \mu}$Jy\,beam$^{-1}$}
\def\mjybm   {${\rm m}$Jy\,beam$^{-1}$}
\def\muJy   {${\rm \mu}$Jy}

\title{A High Resolution Wide-Field Radio Survey of M51}

\author[H. Rampadarath et al]{H. Rampadarath$^{1,2}$\thanks{E-mail: hayden.rampadarath@manchester.ac.uk}\thanks{Currently at: Jodrell Bank Centre for Astrophysics, School of Physics and Astronomy, University of Manchester, Turing Building, Oxford Road, Manchester M13 9PL}, J. S. Morgan$^{1}$, R. Soria$^{1}$, S. J. Tingay$^{1}$,
\newauthor
C. Reynolds$^{1}$, M. K. Argo$^{3}$ and G. Dumas$^{4}$\\
$^{1}$International Centre for Radio Astronomy Research, Curtin University, GPO Box U1987, Perth, WA,
Australia\\
$^{2}$ Department of Physics $\&$ Astronomy, University of Southampton, Highfield, Southampton SO17 1BJ, UK\\
$^{3}$ Jodrell Bank Centre for Astrophysics, School of Physics and Astronomy, University of Manchester, Turing Building, \\ Oxford Road, Manchester M13 9PL\\
$^{4}$ Institut de Radioastronomie Millim\`{e}trique, 300 Rue de la Piscine, F-38406 Saint Martin d'H\`{e}res, France 0000-0002-9833-2948}

\begin{document}
\setlength{\parskip}{0pt}

\date{Accepted ??. Received ??; in original form ??}

\pagerange{\pageref{firstpage}--\pageref{lastpage}} \pubyear{2014}

\maketitle

\label{firstpage}

\begin{abstract}

We present the highest resolution, wide-field radio survey of a nearby face-on star-forming galaxy to date. The multi-phase centre technique is used to survey the entire disk of M51 (77 square arc minutes) at a maximum resolution of 5 milli-arcseconds on a single 8~hr pointing with the European VLBI Network at 18 cm. In total, 7 billion pixels were imaged using 192 phase centres that resulted in the detection of six sources: the Seyfert nucleus, the supernova SN~2011dh, and four background AGNs. Using the wealth of archival data available in the radio (MERLIN and the VLA), optical (\textit{Hubble Space Telescope}) and X-rays (\textit{Chandra}) the properties of the individual sources were investigated in detail. The combined multi-wavelength observations reveal a very complex and puzzling core region that includes a low-luminosity parsec scale core-jet structure typical of AGNs, with evidence for a lateral shift corresponding to 0.27c. Furthermore, there is evidence for a fossil radio hotspot located 1.44~kpc from the Seyfert nucleus that may have resulted from a previous ejection cycle. Our study provides measures of the supernova and star-formation rates that are comparable to independent studies at other wavelengths, and places further limits on the radio and X-ray luminosity evolution of the supernovae SN~1994I, SN~2005cs and SN~2011dh. The radio images of background AGN reveal complex morphologies that are indicative of powerful radio galaxies, and confirmed via the X-ray and optical properties.
\end{abstract}



\section{Introduction}

The formation of stars is a fundamental astrophysical process that takes place in almost all galaxies along the Hubble sequence and as such plays a major role in galaxy evolution. Star formation can occur in isolated areas or over the entire galaxy, resulting in 6-7 orders of magnitude variation in galaxy star formation rates \citep{Kennicutt1998}. Galaxies that are continuously forming stars over an entire galaxy (hereafter star-forming galaxies) typically host rich populations of young stars and star clusters,  HII regions, giant molecular clouds, and supernova remnants (SNRs). Furthermore, it is not uncommon for star-forming galaxies to host low luminosity Seyfert AGN at their cores (e.g. \citealt{CvdH1985}; \citealt{Pellegrini2000}; \citealt{LT09}; \citealt{PT2010}; \citealt{Rampadarathetal2010}; \citealt{Paggi2013}; \citealt{PG13}), suggesting possible connections between star formation and low luminosity AGN activity in these galaxies (e.g. \citealt{Netzer09}).

Nearby star-forming galaxies are ideal laboratories for understanding the nature of supernovae (SNe), SNRs, and the nuclear environment on parsec and sub-parsec scales through the use of wide-field, high resolution radio observations. With multi-epoch radio monitoring, it is possible to construct the star formation and SNe history of these galaxies. This was attempted for the starburst galaxy NGC~253 by \citet{Rampadarathetal2014}, using data spanning 21 years. Similar studies have been conducted of nearby edge-on, star-forming galaxies: M82 \citep{Pedlaretal99, McDonaldetal02, Beswicketal06, Fenechetal2010}; Arp~220 \citep{Rovilosetal05, Lonsdaleetal06, Parraetal07, Batejatetal2012}; Arp~299 \citep{Ulvestad09, Romeroetal11, Bondietal12}; NGC~4945 \citep{LT09}; NGC~253  \citep{UA97, Tingay04, LT06}; and M31 \citep{Morganetal2013}, and nearby face-on, spiral galaxies: M83 \citep{TH94, Cowanetal1994, Maddoxetal2006,Longetal14}; M51 \citep{CvH92, TH94, Maddoxetal07}; NGC 300 \citep{Pannutietal2000}; NGC 7793 \citep{Pannutietal2002}; and NGC 6946 \citep{LG97}. Furthermore, there have been numerous surveys of face-on, spiral galaxies with VLBI, focussing on the nuclear regions (e.g. \citealt{Bietenholzetal2004, GP09, Bontempietal12, PG13, Doietal2013}).

Here, the results of the highest resolution, wide-field radio survey of a nearby, grand design spiral galaxy are presented. The target for this survey is the Whirlpool galaxy, (aka M51a, NGC~5194), that is currently undergoing an interaction with a smaller companion, M51b (aka NGC~5195). At a distance of 8.4 $\pm$ 0.7~Mpc \citep{Vinkoetal12} and with a nearly face-on orientation, M51 is ideal for population and morphology studies at all wavelengths. Like most spiral galaxies M51a is actively forming stars predominantly (but not exclusively) within its spiral arms \citep{Calzettietal05,KS10}, with a slightly enhanced formation of  young, massive star clusters towards M51b \citep{KS10}. Furthermore, in the last 70 years M51a and M51b have hosted  four optically observed SNe: the Type Ia SN~1945A \citep{KS1971}; the Type Ib/c SN~1994I \citep{Puckett1994}; the Type Ib/c SN~2005cs \citep{Muendlein2005}; and the Type IIb SN~2011dh \citep{Griga2011}. In addition, high resolution radio observations of the nuclear region with the VLA reveals the presence of a bidirectional jet, which is associated with a low-luminosity Seyfert 2 AGN, located within the nucleus of M51a, that interacts with the surrounding interstellar medium \citep{CvH92,Bradleyetal2004,Maddoxetal07}. 

The first high-resolution radio survey of M51a was performed by \citet{Maddoxetal07} at 20 and 6~cm, with the VLA. With a resolution of 1.5\arcs $\times$ 1.2\arcs at both wavelengths, \citet{Maddoxetal07} detected 107 compact radio sources, distributed throughout the disk of M51a but predominantly concentrated within the inner 2~kpc nuclear region. \citet{Maddoxetal07} compared the results of the radio observations with multi-wavelength observations and found of the 107 radio sources: 44 radio sources with large HII counterparts; 24 with stellar cluster counterparts; 13 with X-ray counterparts, most probably X-ray binaries or micro-blazars; and six radio sources associated with H$\alpha$ emission, possibly young SNRs with resolved shells and the remaining as either background AGNs or unknown. \citet{Maddoxetal07}  estimated the age for the SNRs to be $\sim$ 2000-3300 yr. 

Through application of the technique of wide-field VLBI, with the European VLBI Network (EVN), we surveyed the full disk of M51  ($11^{\prime}.2 \times 6^{\prime}.9$) for faint, compact radio sources at 18~cm. \Sect~\ref{sec:procedure} describes the observations, wide-field correlation, calibration and data reduction methods that were employed within this study. In \Sect~\ref{sec:DataRed}, the wide-field VLBI imaging, source detection methods, and primary beam correction are discussed. Radio images of the sources detected are also presented. The multi-wavelength and multi-resolution ancillary data (radio, x-rays and optical) used in this study are presented, including data reduction steps, in \Sect~\ref{sec:AncillaryData}. Discussion of the sources detected are given in \Sect~\ref{sec:M51discussion}, including comparisons with the ancillary data and previous results obtained from the literature. Finally, the study is concluded in \Sect~\ref{sec:conclusion}. For power law spectra we adopt the convention that $S\propto \nu^{\alpha}$.

\section{Observations, Correlation and Data Reduction}

\label{sec:procedure}

\subsection{Observations}
\label{sec:obs}

\begin{figure}
\centering
\includegraphics[scale=0.30]{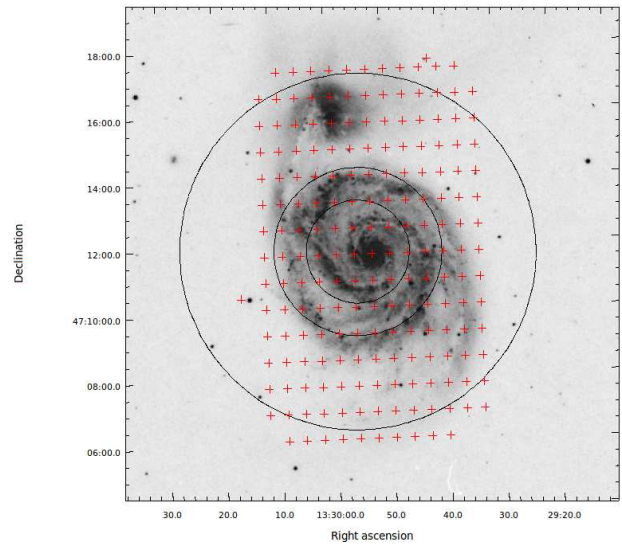} 
\caption{Optical image of M51 from the Space Telescope Science Institute digitised plate archives (image credit: DS9 image server), overlaid with the sensitivity contours of the EVN. The contours are 6.7~$\micro$Jy, 7.5~$\micro$Jy and 13~$\micro$Jy, and are the theoretical sensitivity taking into account the total observing time, bandwidth, the SEFD, effective area, data rate and the primary beam of the individual antennas. The red crosses are the positions of 192 simultaneous multi-phase centres obtained through DiFX used to survey M51 with the EVN.}
\label{fig:EVNsen}
\end{figure}

M51  was observed on 2011-11-7 for 8~hrs (UTC: 01:30 - 09:30) at a wavelength of 18~cm with a single target pointing of the EVN. The experiment was performed using the Effelsberg (Ef), Jodrell Bank 76~m Lovell dish (Jb1), a single Westerbork dish (Wb), Medicina (Mc), Onsala~25~m (On),  Svetloe (Sv), Zelenchukskaya (Zc), Badary (Bd), Urumqi (Ur), and Shanghai (Sh) antennas. During the observation, 5~minute scans of M51 (centred on: RA = 13$^{\mathrm{h}}$29$^{\mathrm{m}}$ 52.698$^{\mathrm{s}}$; Dec = +47$^{\circ}$11$^{\prime}$ 42$^{\prime\prime}$.930 [J2000.0]) were scheduled, alternating with 1 minute scans of a phase reference calibration source, J1332+4722 (RA = 13$^{\mathrm{h}}$ 32$^{\mathrm{m}}$ 45.246424$^{\mathrm{s}}$; Dec = +47$^{\circ}$ 22$^{\prime}$ 22$^{\prime\prime}$.667700 [J2000.0]) located 31$^{\prime}$ from M51.

\begin{figure}
\centering
\includegraphics[scale=0.22]{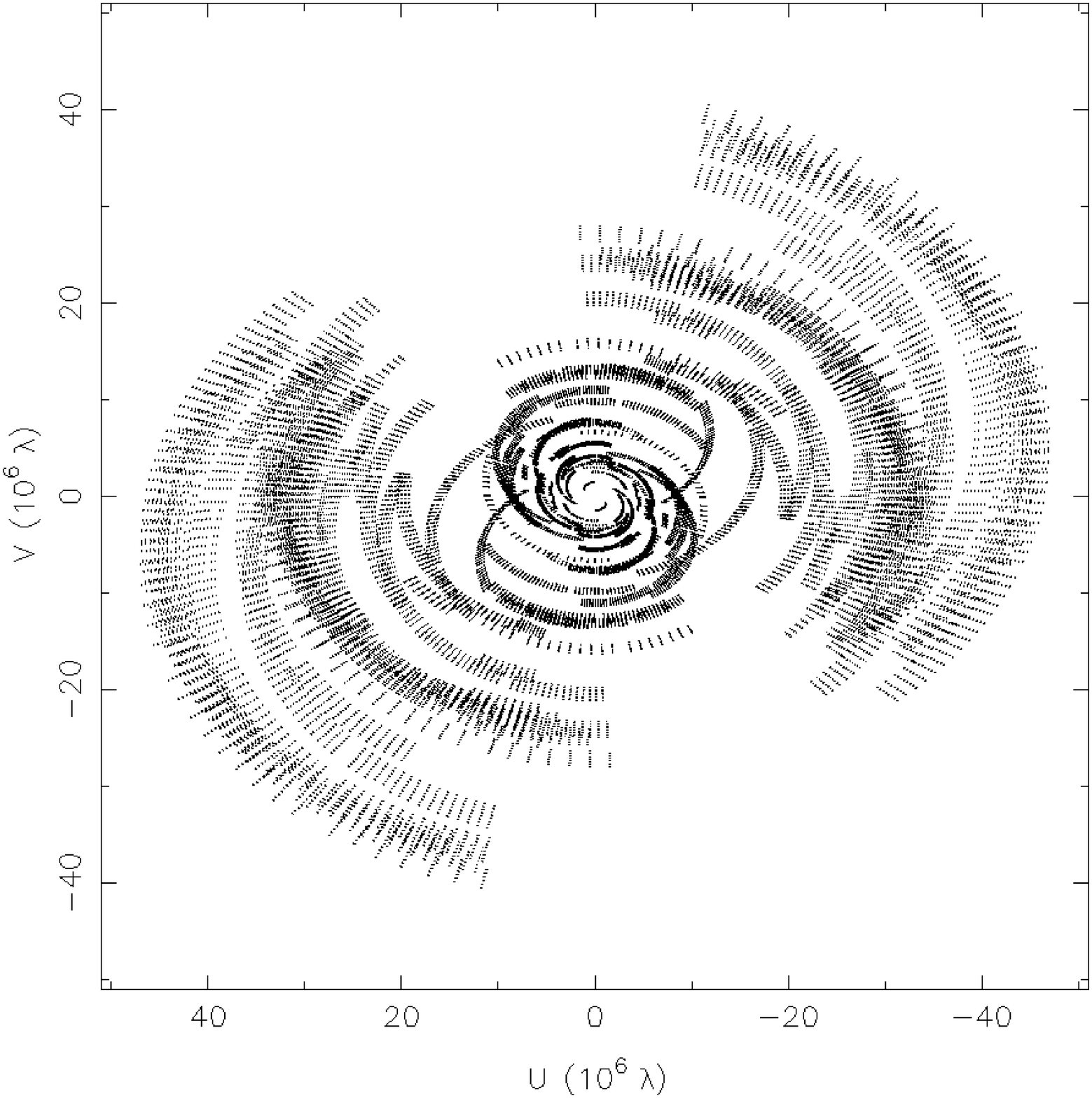} 
\caption{The $(u,v)$ coverage of the 10-station EVN observation at 18~cm (One point per sub-band per target scan, for each baseline only).}
\label{fig:M51uvcov}
\end{figure}

The VLBI data were recorded in 8$\times$16~MHz sub-bands, dual polarisation with 2-bit data sampling at the Nyquist rate, resulting in a total data rate of 1024 Mbps, producing a 1$\sigma$ sensitivity (at the pointing centre) of 6$\micro$Jy. \Fig~\ref{fig:EVNsen} shows an optical image of M51 overlaid with the sensitivity contours of the EVN observation, taking into account the different primary beam of each antenna.
The angular size of M51 ($11^{\prime}.2 \times 6^{\prime}.9$) allows most of the galaxy to lie within the primary beam of the 100~m Effelsberg antenna ($\sim7^{\prime}.5$).
\Fig~\ref{fig:M51uvcov} shows the $(u,v)$ coverage of the EVN observation. The maximum projected baseline is approximately 45~M$\lambda$ (Jb1-Sh), while the shortest projected baseline (Ef-Wb) is 1.5~M$\lambda$, hence structures with an angular size larger than 150 mas (5.7~pc at the distance of M51) are resolved-out (i.e. structures with intensity variations on larger angular scales are not detected).

\subsection{Correlation}

To survey the entire disk of M51 for radio sources with the full capabilities at VLBI resolution requires application of the technique of multi-phase centre correlation \citep{Delleretal11,Morganetal11}. This was achieved through the use of the DiFX software correlator \citep{Delleretal07, Delleretal11}, at the Max-Planck Institue for Radio Astronomy, Bonn, Germany, as at the time of this experiment the JIVE software correlator (SFXC) was not offering multi-phase centre correlation\footnote{The SFXC is currently capable of correlating multiple phase centres and this is offered as a standard EVN mode.}. 
Prior to correlation, the data were shipped via courier from each station to Bonn, where the correlation was performed with a high performance computer cluster  comprising 60 computer nodes (8 cores each).

The correlation was initially performed with sufficient frequency and time resolution (488~Hz and 1~ms) to restrict amplitude losses below $5\%$ at a radius of 6\arcm on the longest baseline due to bandwidth and time smearing\footnote{\url{http://www.atnf.csiro.au/people/Emil.Lenc/Calculators/wfcalc.php}}. Then using the simultaneous multi-phase correlation method, the field was separated into 192 individual phase centres, allowing the area around each phase center to be imaged separately. The locations of the phase centres are shown in \Fig~\ref{fig:EVNsen}. 

To minimise cumulative amplitude errors to less than 5$\%$ due to non-coplanar baselines, bandwidth, and time-average smearing on the longest baseline per phase centre, each sub-band (of 16~MHz bandwidth) was averaged to 128 channels ($\delta\nu$ = 125~kHz) and time resolution of 1 second, resulting in a data size of 41~GB per phase centre. The correlated data were transferred to the iVEC Petabyte Data Store\footnote{\url{http://www.ivec.org/}} located at Curtin University, Perth, Western Australia, using the network data transfer protocol gridFTP\footnote{\url{http://www.globus.org/toolkit/docs/latest-stable/gridftp/}}.

\subsection{Data Reduction}

\subsubsection{Data Preparation}

During correlation, the clock at the Badary antenna experienced a glitch that caused it to skip ahead by 1s. This resulted in a separation of the final correlated data per phase centre, at the time of the clock skip, into two individual FITS files. Concatenation of the two correlated files per phase centre to form the individual visibility FITS via the \textsc{aips} task \textsc{fitld} takes $\sim$2 hours on a standard iCore 7 desktop computer. Extending this process to 192 phase centres would require more than 300 hrs of computing time, which is almost tripled when considering read and write time of the FITS data to disk. 

In order to concatenate, calibrate and image the 192 phase centres in a time that is suitable for this project, the Curtin ATNF VLBI Experiment (\textsc{cave}), which is a high performance computing cluster located at the Australia Telescope Compact Array (ATCA), Narrabri, NSW was used. \textsc{cave} comprises 14 computer nodes, where the nodes are quad-core 2.7~GHz AMD Sunfire x86 machines, with 16~GB memory and 400~GB disk capacity each. The main disk space (8~TB $\times$ 3) is located on a single centralised node (hereafter, \textsc{cave-store}). The cluster is owned by Curtin University and operated by the Australia Telescope National Facility (ATNF). To use the multiple nodes of \textsc{cave} effectively for the concatenation, we made use of the \textsc{ParselTongue} scripting language \citep{Kettenisetal06}.  \textsc{ParselTongue} allows the execution of multiple \textsc{aips} tasks simultaneously on a number of remote computers via the \textsc{paralleltask} module. 


The correlated datasets were copied from the iVEC Petabyte Data Store to the 8~TB data storage on \textsc{cave-store}. Before starting the concatenation process, the \textsc{ParselTongue} server is started on the individual nodes\footnote{See the \textsc{ParselTongue} online documentation  \url{http://www.jive.nl/jivewiki/doku.php}}. Executing the \textsc{ParselTongue} script on \textsc{cave-store} begins by copying the datasets to the individual nodes, such that the datasets for the N$^{\mathrm{th}}$ phase centre is sent to the N$^{\mathrm{th}}$ node. After the data are copied, \textsc{fitld} loads and concatenates both datasets to \textsc{aips} and the result written to disk (via \textsc{fittp}), copied to \textsc{cave-store} and eventually is copied to iVEC. Note, the data processing on the individual nodes is executed in parallel. 

Due to the limited storage space and network connection between iVEC and \textsc{cave-store}, the datasets were separated into $\sim$1~TB groups ($\sim$ 25-30 phase centres) for processing. The time taken to process $\sim$1~TB of data on two nodes on \textsc{cave} was 14 hrs\footnote{The processing time per 1~TB group can be speeded up by using multiple nodes for the data processing i.e. three nodes would reduce the time to 9~hrs, four nodes to 7 hrs etc. However, \textsc{cave}, is a shared cluster and only two nodes were fully available for this project at any time.}, resulting in a total time of $\sim$87 hrs to concatenate the 192 phase centres.


\subsubsection{Calibration}

\begin{figure}
\centering
\includegraphics[scale=0.3,angle=-90]{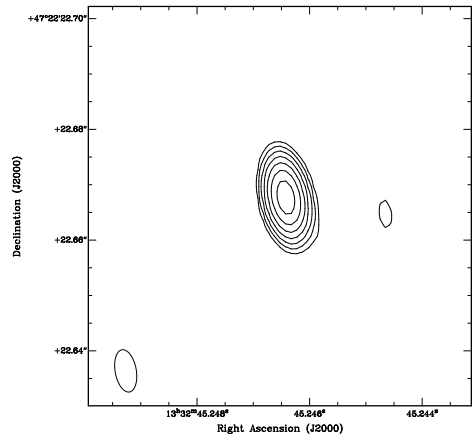} 
\caption{Contour map of the calibrator J1332+4722, with a dynamic range of $\sim$ 2500 between peak flux density and the RMS  for regions in the image away from the inner side-lobes around the source. The contours begin at $\pm1\%$ of the peak (226.02~mJy/beam), and increase in increments of $2^{n}\%$ of the peak where $n$ are integers (1...7). }
\label{fig:calib}
\end{figure}

The data reduction and calibration were performed using the data reduction package \textsc{aips}. Prior to calibration, flagging of data during times at which each of the antennas were known to be slewing and time ranges that contained known bad data was carried out via application of flag files and information provided in the observing logs. Corrections to the amplitudes in cross-correlation spectra from the auto-correlations were obtained via \textsc{accor}. 
The amplitude calibration was obtained through application of antenna system temperatures (in Kelvin) measured during the observations, along with the gain (in Janskys per Kelvin) for each antenna. Following the amplitude calibration, a single value of the delay (i.e. phase against frequency) was determined for each antenna, polarization, and 16~MHz sub-band using two minutes of data on the bright radio source 3C~345 (\textsc{aips} task \textsc{fring}). Global fringe-fitting solutions were determined for J1332+4722 (\textsc{aips} task \textsc{fring}) with a one minute solution interval, finding independent solutions for each of the 16 MHz sub-bands. The delay and phase solutions were examined and, following editing of bad solutions, applied to J1332+4722.
The J1332+4722 data were then exported to \textsc{difmap} \citep{S97}, where the data were vector-averaged over 30~s, flagged of bad data, and imaged using standard imaging techniques (deconvolution and self-calibration of both phase and amplitude). The resulting image of J1332+4722 is shown in \fig~\ref{fig:calib}. The image shows no evidence for extended structure on these baselines at this frequency to a dynamic range of 2500, for regions in the image away from the inner side-lobes around the source. The final calibration solutions (phase and amplitude) from J1332+4722 were exported via the \textsc{difmap} extension, \textsc{cordump}\footnote{\url{http://www.atnf.csiro.au/people/Emil.Lenc/tools/Tools/Cordump.html}} \citep{LT09} to an \textsc{aips}-compatible solution table. The solution table was then transferred to \textsc{aips} and applied to all sources in the dataset. The final calibrated J1332+4722 data were also used to derive a bandpass calibration via the \textsc{aips} task \textsc{bpass}. The edge channels of each band were flagged (5 channels from both the lower and upper edge of each 128 channel band). The final calibration solutions were then applied for a first imaging pass for each phase centre.


\section{Imaging, Source Detection and Results}

\label{sec:DataRed}

\subsection{Wide-field VLBI Imaging}
\label{sec:WFImaging}

Prior to imaging, data affected by RFI were manually flagged via the \textsc{aips} tasks \textsc{spflg} and \textsc{editr}. 
Data received by the Mc antenna were found to be corrupted by RFI. Initial attempts at selective RFI removal were found to be excessively time consuming and the entire antenna was flagged, improving the final sensitivity by $\sim10\%$. Furthermore, data with $u$ coordinate $<50$~k$\lambda$ which are susceptible to RFI due to the low fringe rate were also flagged. The flag tables along with the final calibration and bandpass tables were copied and applied to all phase centres via \textsc{split}.

\begin{figure}
\centering
\includegraphics[scale=0.35]{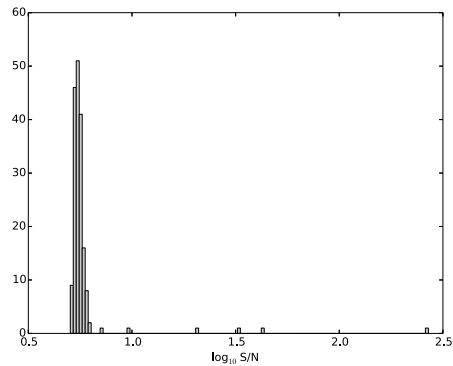} 
\caption{ The distribution of the peak signal to noise ratio (S/N) for the 192 phase centres, where the x-axis is plotted in logscale.}
\label{fig:SNRhist}
\end{figure}

Initial imaging was restricted to the four European antennas (Ef, Jb1, Wb and On) resulting in a maximum baseline of 7.5~M$\lambda$ (resolution of  27.6 mas). This allowed a reduction in the number of channels to 32 per sub-band and increased time averaging to 5 seconds, while keeping the total amplitude loss due to time and bandwidth smearing $<5\%$. Moreover, the decreased resolution lowered the number of pixels required to image a radius of 21\arcs per phase centre and, therefore, the data size per phase centre, thus reducing the computation required to image the phase centres. Dirty maps of all phase centres were produced via the \textsc{aips} task \textsc{imagr}, using parameters: cellsize = 2.7~mas; image size = 16384 $\times$ 16384 pixels; and natural weighting. Finally, 512 pixels were removed from the image edges, to reduce aliasing effects. This resulted in image sizes covering 41$^{\prime\prime} \times 41^{\prime\prime}$ per phase centre.
The \textsc{aips} task \textsc{imean} was then used to obtain the RMS noise levels across the entire image, allowing comparisons with the peak pixel brightness. 


The distribution of the peak S/N for the 192 phase centres is displayed in \Fig~\ref{fig:SNRhist}. Six phase centres were found to have a peak S/N $>7\sigma$, with the remaining phase centres distributed between 5$\sigma$ and 6.15$\sigma$. While the six peak S/N are indeed real detections (and are discussed in \Sect~\ref{sec:M51discussion}), it is possible that there are real detections with S/N $<7\sigma$.


\subsection{Source Identification}

\subsubsection{Defining the detection threshold}
\label{sec:threshold}

Morgan et al. (2013) addressed the problem of determining the likelihood that a bright pixel in a large VLBI image is a source (rather than a Gaussian noise spike). Consider a VLBI image of pixel size $n_{\mathrm{pixels}}\,\times\,n_{\mathrm{pixels}}$. Assuming that the noise in the image is Gaussian distributed and is free from RFI or emission from real sources, the probability that a pixel exists with S/N greater than some threshold, $S_{m}$, can be shown to be

\begin{equation}
P(\mathrm{S/N}\geq S_{m}) = 1- \mathrm{erf}\left( \dfrac{S_{m}}{\sqrt{2}}\right)
\label{eq:erfc}
\end{equation}

where erf(x) is the error function associated with the Gaussian distribution.
\Eqn~\ref{eq:erfc} is the probability that an image pixel of S/N is above a given threshold, assuming the noise is Gaussian distributed. Thus, pixels found above this threshold can be considered as being from real sources. The reciprocal of \eqn~\ref{eq:erfc} is therefore the equivalent frequency of occurrence, $N_{\mathrm{freq}}$, which is a measure of the total number of image pixels such that a single pixel will have S/N equivalent to $S_{m}$. 

\begin{figure}
\centering
\includegraphics[scale=0.35]{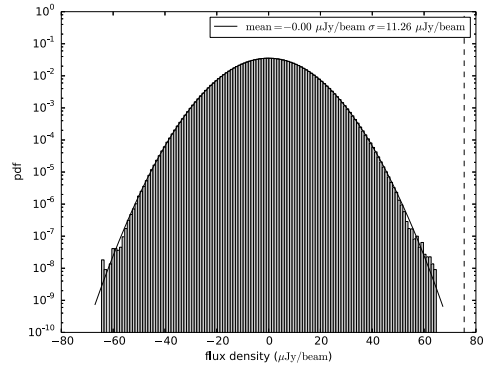} 
\caption{Histogram showing the pixel distribution of a single 15360 pixels $\times$ 15360 pixels wide-field VLBI image, with the y axis in log scale. The solid line represents a least squares Gaussian fit to the distribution, with the mean and standard deviation given inset. The dashed lines plot the location of the 6.7$\sigma$ detection threshold.}
\label{fig:pixel_hist}
\end{figure}

\Fig~\ref{fig:pixel_hist} plots the pixel distribution of a single 15360 pixels $\times$ 15360 pixels wide-field VLBI image, representing a blank region of the sky, 5 arcminutes from the pointing centre. The field covered by this image was found to be devoid of radio sources in the FIRST  \citep{FIRST95} and \citet{Maddoxetal07} radio catalogues\footnote{See \Sect~\ref{sec:frac} for a description of these radio source catalogues}, and lacked pixels with a flux density above the Gaussian noise threshold. The distribution follows a Gaussian distribution (the solid line), suggesting that the pixels of the wide-field VLBI images are Gaussian distributed\footnote{Note that this is only valid if the data are free from real sources and RFI (achievable through careful flagging). This is shown empirically in \citet{Morganetal2013} and \Fig~\ref{fig:pixel_hist}}. 

Since multiple phase centres provide a method to image a large field through $N_{\mathrm{field}}$ smaller fields, the total number of pixels is $N_{\mathrm{field}}\,\times\,n_{\mathrm{pixels}}\,\times\,n_{\mathrm{pixels}}$.
By assuming the pixel brightness distribution in the wide-field images is Gaussian distributed, from \Eqn~\ref{eq:erfc}, one 5.8$\sigma$ pixel would be expected on average in every 15360 pixel $\times$ 15360 pixel image and one 6.7$\sigma$ pixel in 192 images, assuming no astronomical sources. This is in agreement with \Fig~\ref{fig:pixel_hist}, which clearly shows there are no pixels on either ends of the distribution beyond 6.7$\sigma$. Thus, any peak flux density with S/N $\geq6.7\sigma$ is more likely to be a source than a noise spike (assuming real sources are the only thing that could make the distribution non-Gaussian).

\subsubsection{Using Prior Catalogues to Search for Faint Sources}
\label{sec:LRmethod}

The detection threshold calculated in \sect~\ref{sec:threshold}, assumes that the image
 pixels are independent. This is  not the case, as pixels within an area defined by the synthesised beam may be correlated. With a cellsize of 2.7~mas and resolution of 27.6~mas, gives $\sim$100 pixels per synthesized beam, reducing the threshold to 5.1$\sigma$ in a single 15360 pixel $\times$ 15360 pixel image and one 6.0$\sigma$ pixel in 192 images.

However, there is a possibility that real sources exist in the wide-field images below this limit and are treated as a noise-spike. To determine this probability the likelihood ratio (LR) method described by \citet{Morganetal2013} is applied. This method calculates the likelihood that each pixel in a VLBI image is associated with a catalogued source of known position, size and flux density (the alternative hypothesis, $H_{1}$), or simply a noise spike (the null hypothesis, $H_{0}$). This calculation is made assuming the noise in each image is Gaussian distributed and the RMS of the noise does not vary across an individual image. This method generates LR images corresponding to the input VLBI images, where the pixel values correspond to the likelihood that the pixel is related to a catalogued source (see \Sect~2.5 of \citealt{Morganetal2013} for a detailed description of this method). 

We use this method to search for VLBI counterparts of sources detected in M51 by the VLA FIRST survey \citep{FIRST95} and \citet{Maddoxetal07} using the VLA in A configuration. In their analysis, \citet{Morganetal2013} found that the results of the LR test did not give a significant advantage over simply detecting bright pixels due to the poor constraints on the position given by the lower resolution catalogues they used. Given the better positional accuracy provided by the VLA FIRST \citep{FIRST95} and the \citet{Maddoxetal07} surveys, it was deemed worthwhile to attempt to use this method.

However, similar to \citet{Morganetal2013} we found that the LR of a bright pixel was more dependent on the VLBI pixel S/N (i.e. p($H_{0}$)), than the positional constraints given by the lower resolution catalogues (see Appendix A for further details on this method and the results). As such no significant advantage is given by the LR method over simply detecting the brightest pixels. Thus, it is difficult to determine whether pixels with S/N$<6.7\sigma$  are indeed real detections, and are henceforth considered as resulting from noise.

\subsection{Primary Beam Correction}
\label{sec:m51pbcor}

Primary beam corrections in VLBI observations are uncommon and to-date have been attempted for a subset of EVN antennas \citep{Morganetal11} and the VLBA \citep{Middelbergetal11,Morganetal2013,Middelbergetal2013}. Unlike interferometers such as the JVLA where the primary beam correction is performed in the image plane, the approach taken by \citet{Middelbergetal11} determines the primary beam attenuation through calculating the visibility gains. This requires the assumption that the correction is constant across the entire image. 

Primary beam corrections were applied to the phase centres containing pixels above 6.7$\sigma$, following the method of \citet{Middelbergetal11} and \citet{Morganetal2013}. Very limited information is available on the primary beams of EVN dishes and so our primary beam model is necessarily simplistic, assuming a rotationally symmetric Bessel function with a FWHM calculated from the published diameter of the dish\footnote{see \url{http://www.evlbi.org/user_guide/EVNstatus.txt}} for both polarisations. For Effelsberg only, we used the FWHM quoted in \footnote{\url{https://eff100mwiki.mpifr-bonn.mpg.de/doku.php?id=information_for_astronomers:rx:p200mm}} scaled relative to the reference frequency quoted therein. For all dishes the primary beam was calculated using the central frequency of each subband, for the duration of the observation, using the measured system temperatures for each antenna and then applied to the visibilities.

It should be noted that the primary beam correction did not improve the S/N of the detected sources. This is evidence of fairly significant errors in the model, since a correct model would result in optimal weighting of each baseline. For the EVN, unlike the VLBA array, the difference in antenna sizes (from 100 m to 25 m) results in very different primary beam responses. Moreover, during the course of the observations the different physical characteristics of the individual elements produce varied effects on the overall primary beam response of the array\footnote{Note:- this is also true for the VLBA}. Furthermore, the exact characteristics of each antenna in the EVN (except for Ef to a certain degree) are unknown\footnote{There are current efforts being made to measure the primary beams of the EVN and MERLIN antennas, (R. Beswick, private communication)} and a number of assumptions were required (e.g uniform illumination, symmetric primary beam, and a perfect parabola defined by a Bessel function), which would no doubt affect the final correction. While there are errors in the final primary beam-corrected flux values, they are considerably less than not applying the primary beam correction. Thus, there is no prospect of discovering new sources by re-imaging with a primary beam correction applied to the visibilities. Nonetheless, in estimating the flux of the sources, the overall primary beam correction is taken into account and the source fluxes are adjusted accordingly.

For radio sources within the primary beam of the Ef antenna (radius = 3.8\arcm) we estimate the residual amplitude error  due to the primary beam is on the order of 10$\%$. While for sources beyond the HPBW of the Ef primary beam, an error of 20$\%$ is estimated. This term is added in quadrature with the other sources of error which is listed in \Tab~\ref{tab:VLBI_Sources}.

%

\subsection{Flux Density and Morphology}
\label{sec:M51Fluxes}
\begin{figure*}\scriptsize
\centering
\subfloat[J132952+471142]{\includegraphics[scale=0.3]{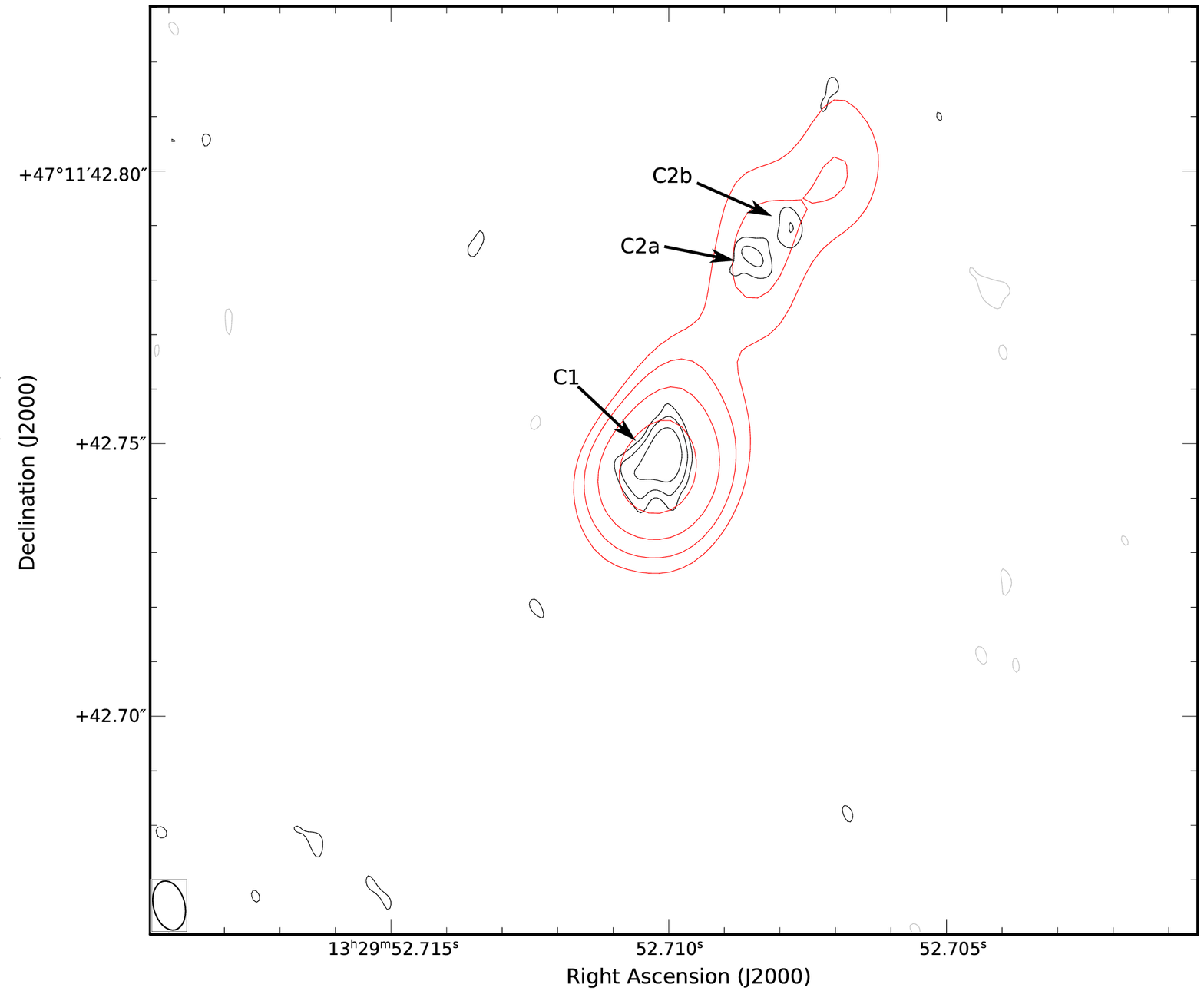}}
\subfloat[J133005+471035]{\includegraphics[scale=0.3]{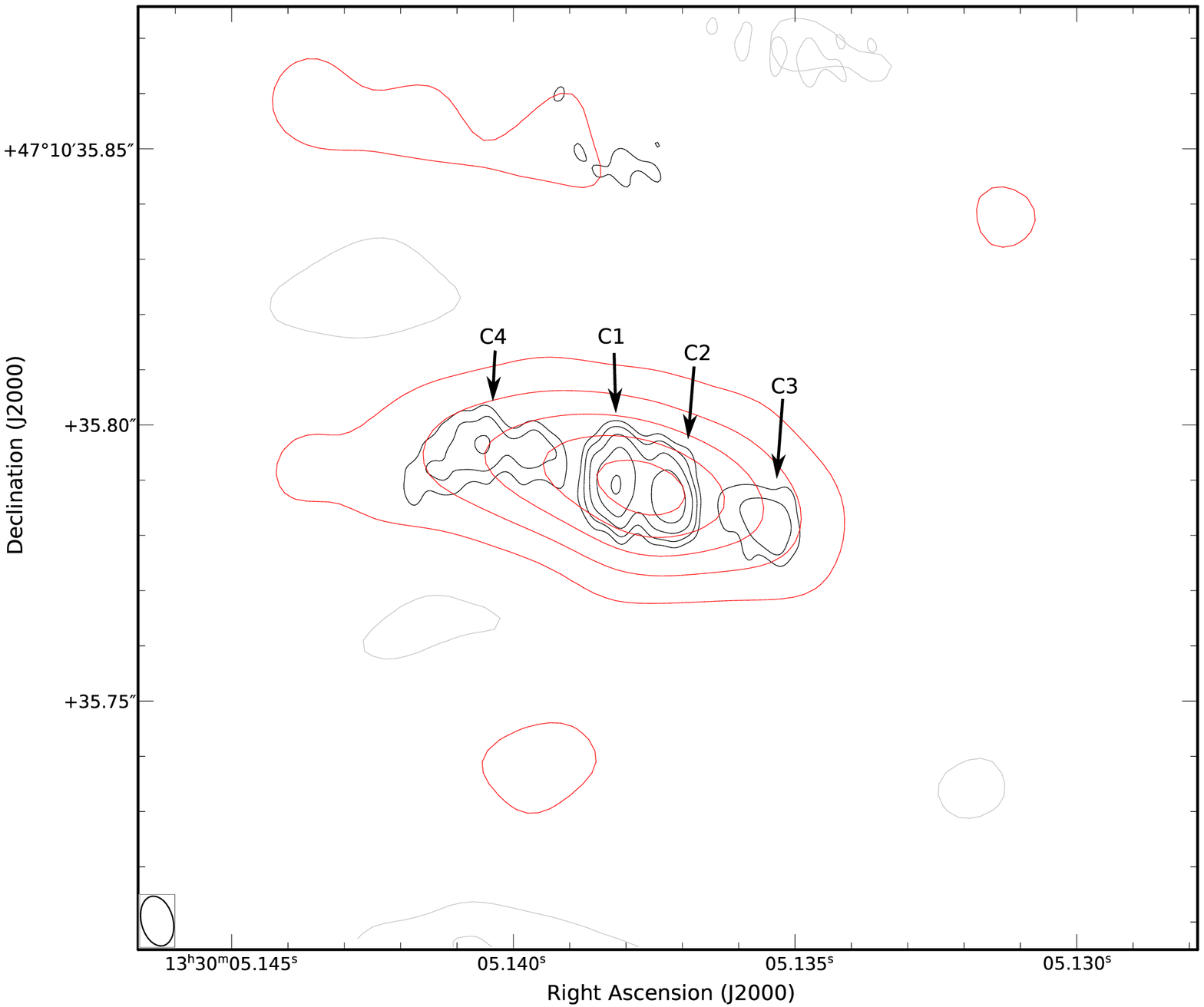}}\\
\subfloat[J133005+471010]{\includegraphics[scale=0.3]{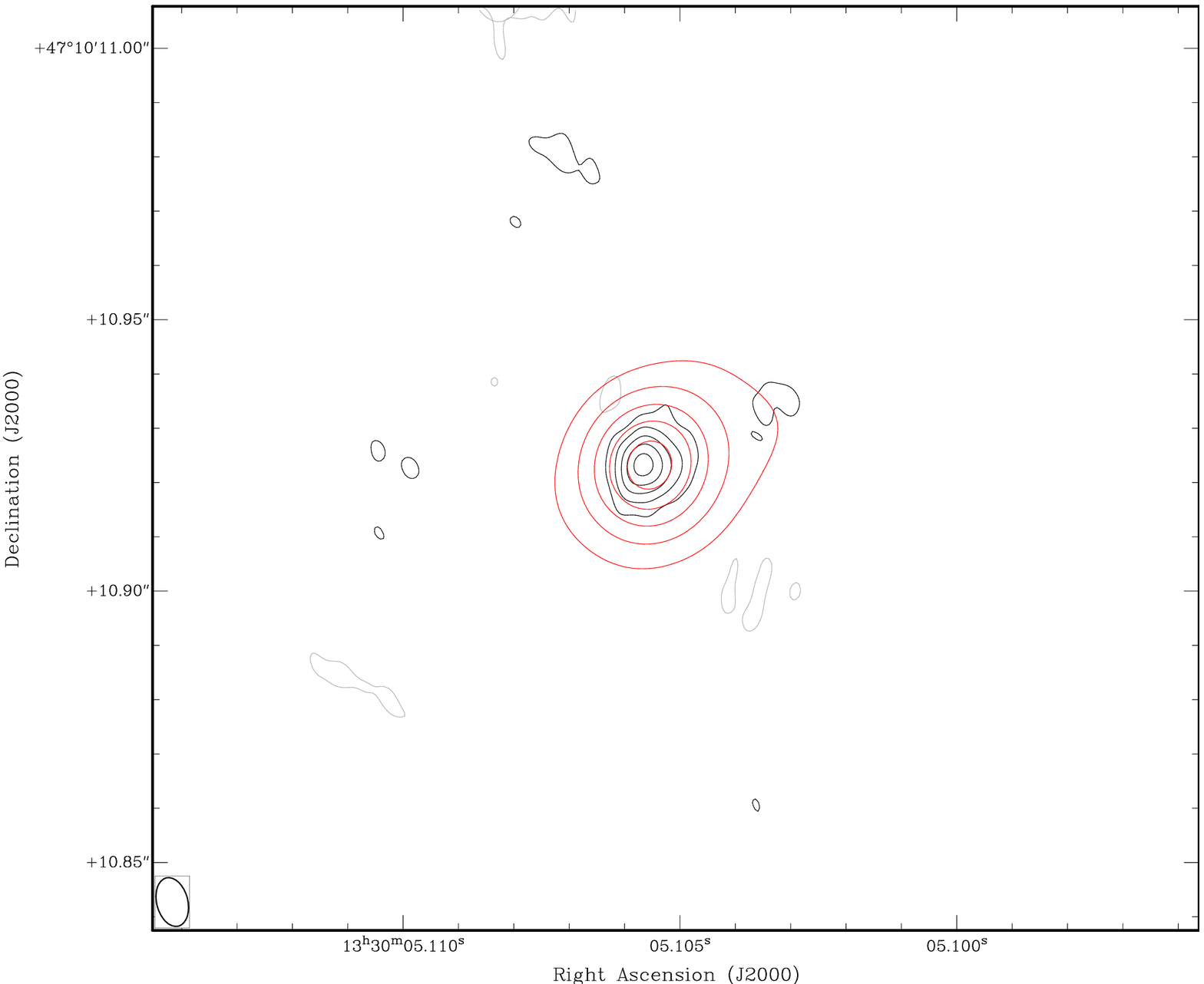}}
\subfloat[J133011+471041]{\includegraphics[scale=0.3]{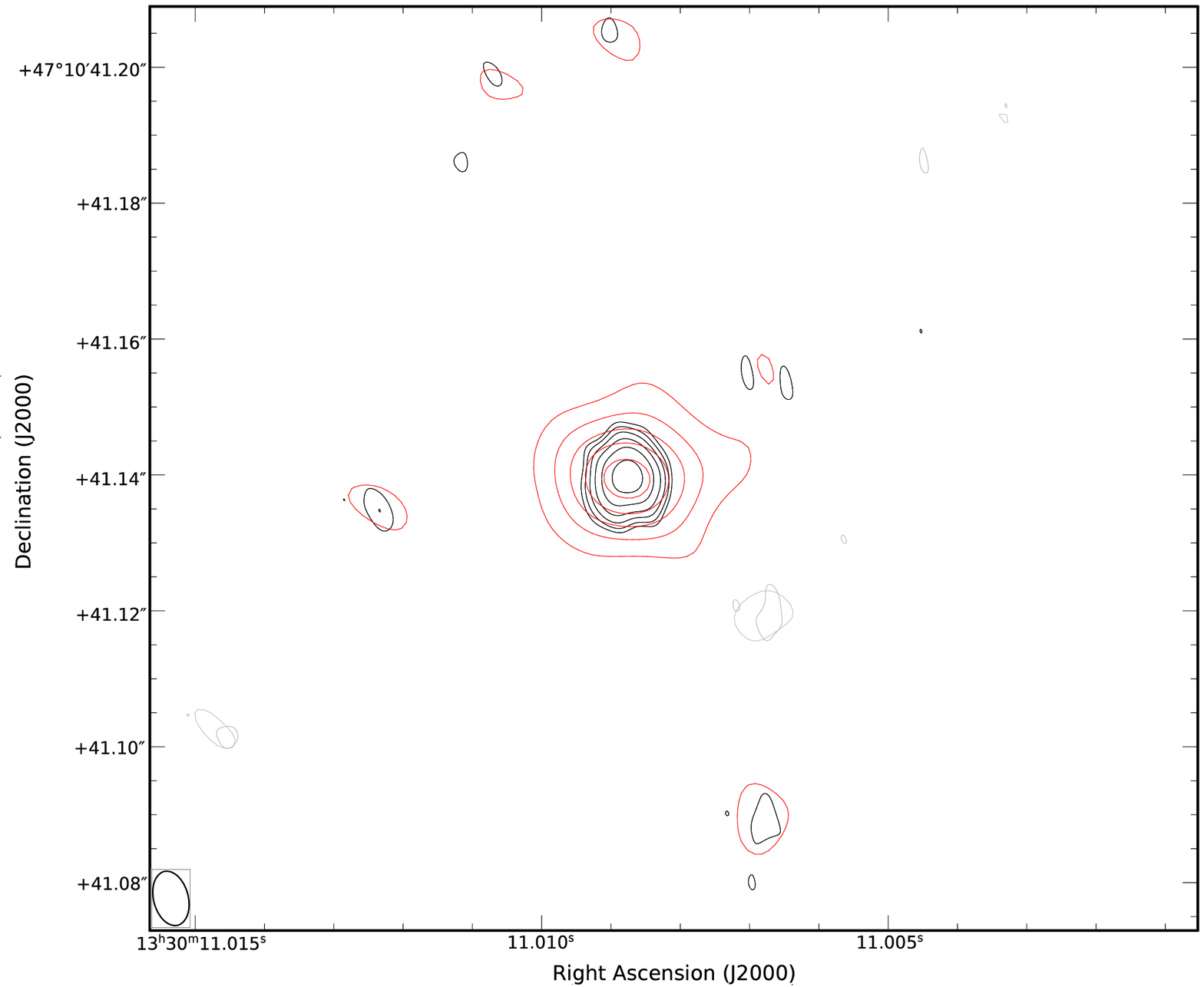}}\\
\subfloat[J132932+471123]{\includegraphics[scale=0.3]{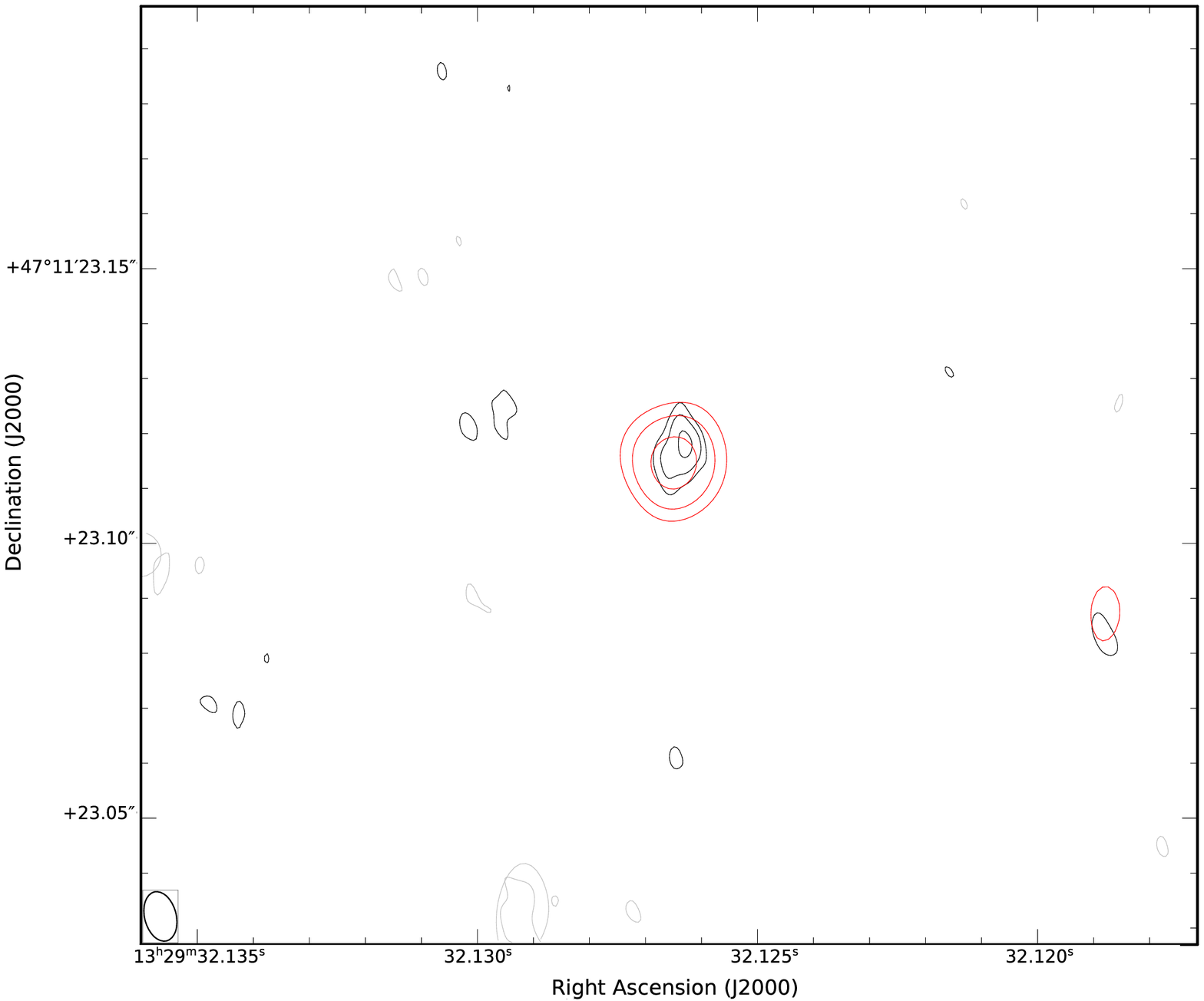}} 
\subfloat[J133016+471024]{\includegraphics[scale=0.3]{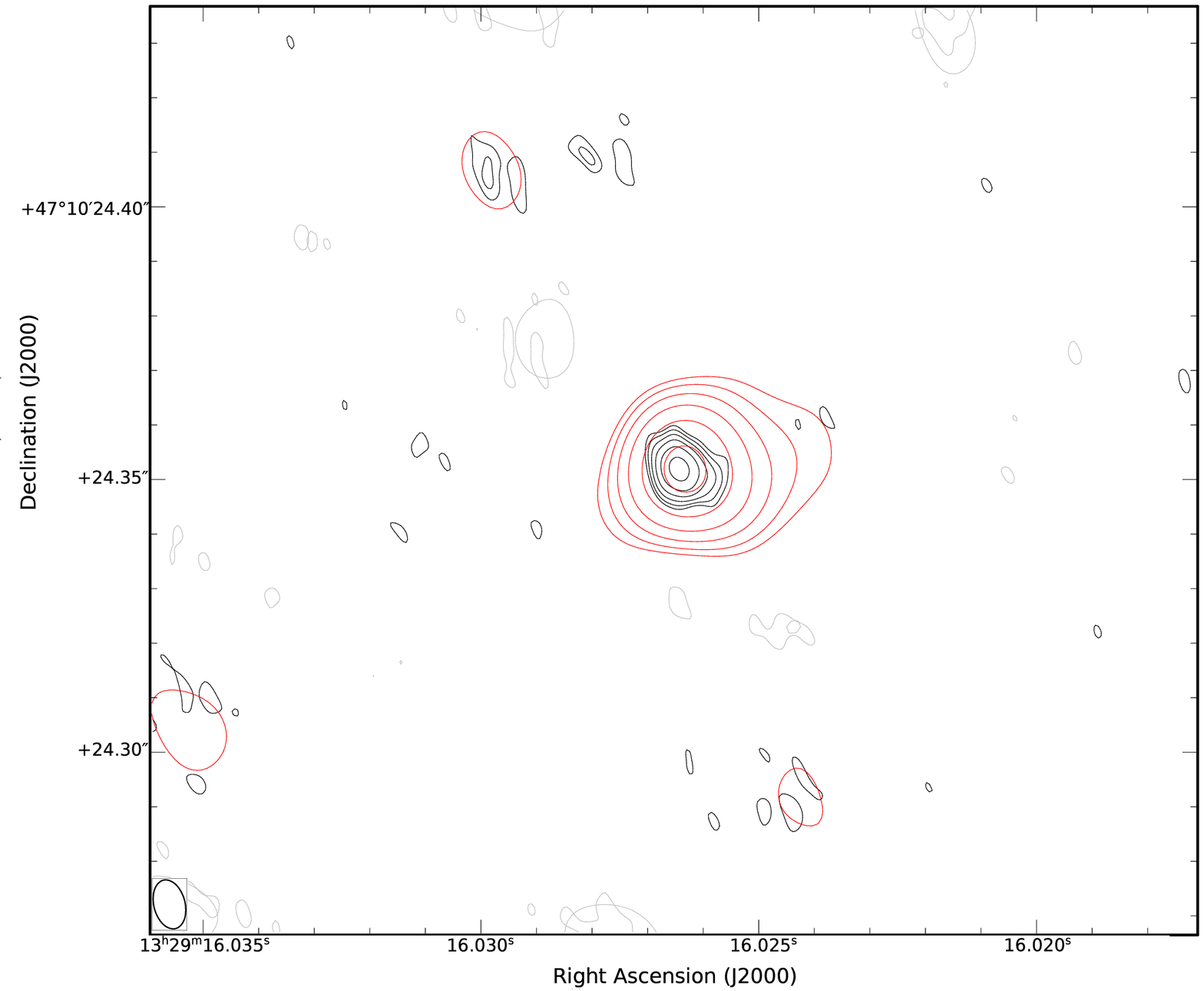}}
\caption[Sources detected with the EVN at 18 cm in M51]{\scriptsize Sources detected with the full array of the EVN at 18~cm. The black contours are full resolution images with robust weighting 4, with a beam size = 9.86~mas $\times$ 5.92 mas and  position angle = 14.3\deg.  The lowest contours (for all but J133005+471010) are at $\pm3\sigma$, and increase in multiples of $\sqrt{2}$, where $\sigma$ is the RMS of the images found via \textsc{imean}: 9.1~\mujybm ~(J132952+471142); 17.4~\mujybm (J133005+471035); 18.3~\mujybm ~(J133011+471041); 18.0~\mujybm ~(J132932+471123);  18.6~\mujybm ~(J133016+471024). The contours of J133005+471010 begins at $\pm10\%$ the peak flux density  (3.26~mJy/beam), and increase in increments of $20\%$ the peak flux. The sources were also imaged with a $(u,v)$ range taper of 10 M$\lambda$ (red contours), with a resulting a beam size = 25.3~mas $\times$ 21.2 mas and  position angle = -4.9\deg.  The lowest contours are $\pm3\sigma$, and increase in multiples of $\sqrt{2}$ for; J132952+471142 ($\sigma$=10.5~\mujybm), J132932+471123 ($\sigma$=21.8~\mujybm) and J133016+471024 ($\sigma$=24.4~\mujybm). While the remaining sources the contours begin at $\pm10\%$ the peak flux density (J133005+471035, 506.6~\mujybm; J133005+471010, 4.27~\mjybm; and J133011+471041, 552.6~\mujybm), and increase in increments of $20\%$ the peak flux density .}
\label{fig:VLBI}
\end{figure*}

\begin{figure}
\centering
\includegraphics[scale=0.35,clip=false,trim=0.5cm 0.1cm 0cm 0cm]{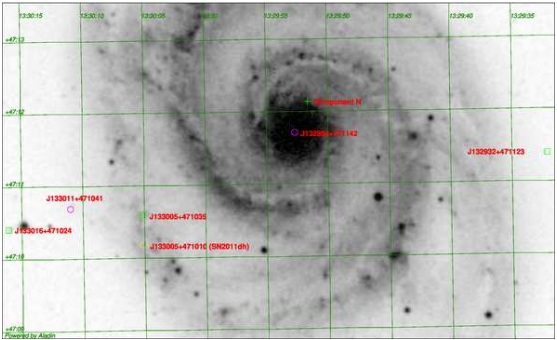} 
\caption[Image showing the location of the six sources detected with the EVN in M51.]{Location of the six sources detected with the EVN shown in \Fig~\ref{fig:VLBI} and listed in \Tab~\ref{tab:VLBI_Sources}. Flat spectrum sources ($-0.5\,\leq\,\alpha\,\leq\,0.5$) are the pink circles and steep spectrum sources $\alpha\,\geq\,0.5$ are the green squares. The yellow elliptical source is the supernova SN~2011dh, that exploded 159 days prior to our EVN observation (see \Sect~5.5.3). The green cross is a steep spectrum source detected with MERLIN at 18~cm (designated Component N) but not with the EVN at 18~cm (see \Sect~5.5.2).}
\label{fig:src_loc}
\end{figure}

In total six sources were found using the \textsc{blobcat} source extractor \citep{Halesetal12} with flux densities above 6.7$\sigma$, which are listed in \Tab~\ref{tab:VLBI_Sources}, contour maps and locations of the sources in the field are displayed in \Figs~\ref{fig:VLBI} and \ref{fig:src_loc}, respectively. 
Following the correction for the amplitude response of the primary beam, the positions of the six sources were shifted to the image centre via \textsc{uvfix}, averaged in frequency, and reimaged. 

\textsc{clean}ed images were obtained using data from the full array via the \textsc{aips} task \textsc{imagr}. The images were made with cellsize = 0.5~mas, imsize = 4096 pixels $\times$ 4096 pixels, and 100 clean iterations. Briggs robust parameter 4 was used to weight the images as it was found to produce an acceptable compromise between sensitivity and resolution. The final images are displayed in \Fig~\ref{fig:VLBI}.

Measurements of flux density and morphology were obtained by fitting a single 2D Gaussian component via \textsc{imfit}, with independent measures of flux density from \textsc{blobcat} and source sizes from fitting a circular Gaussian to the visibilities via \textsc{modelfit} in \textsc{difmap} \citep{S97}. The measurements are listed in \Tab~\ref{tab:VLBI_Sources}, with any cross identification with the  FIRST \citep{FIRST95} and \citet{Maddoxetal07} surveys. The different measurements of flux density and source size were found to be consistent, so only the \textsc{imfit} values of  flux density and source size are included in \tab~\ref{tab:VLBI_Sources}. 

Furthermore, to search for possible emission associated with extended structures, lower resolution maps with the full array were made with a $(u,v)$ range taper of 10 M$\lambda$ for all sources and are included in \Fig~\ref{fig:VLBI} (the red contours). Each source will be discussed in detail in \Sect~5.5.

\begin{center}
\begin{table*}\footnotesize
\caption{Sources Detected in M51 with the EVN at 18 cm. \emph{Columns:} (1) Source name; (2) $\&$ (3) Source position; (4) Angular distance to the centre of M51. At 8.4~Mpc, 1\arcm $\approx$ 2.44 kpc; (5) Signal to noise ratio of the source in the wide-field source detection maps; (6) Flux density measured from \textsc{imfit} using a single 2-D Gaussian component; (7) $\&$  (8) Source sizes measured via \textsc{imfit}; (9) Cross identification  with the \citet{Maddoxetal07} survey; (10) Detected by FIRST (Y), or not (N).}
{\renewcommand{\tabcolsep}{3.5pt}
\begin{tabular}{cccccccccc}
\hline \hline
&\multicolumn{ 2}{c}{Position (J2000.0)} & & &  & \multicolumn{ 2}{c}{IMFIT Size}  &  &  \\ 
\cline{2-3}  \cline{7-8} 
&RA & Dec &  $\Theta$& & F$_{\mathrm{IMFIT}}$  & Major & Minor &  \multicolumn{ 2}{c}{Identification} \\ 
\cline{9-10}
Source&13$^{h}$ & 47\deg & (\arcm)  & S/N & ($\mu$Jy)  & (mas) & (mas)   & Maddox & FIRST \\ 
(1) &(2) & (3) & (4)  & (5) & (6)  & (7) & (8)   & (9) & (10) \\ 
\hline
J132952+471142$^{\diamond}$ &29$^{m}$52.7102(3)$^{s}$ & 11\arcm 42.7465(5)\arcs &  0 & 9.7  & 140.1$\pm$16.0  & 8.8$\pm$1.1 & 1.9 $\pm$0.9  & 53 & Y \\ 
J133005+471035&30$^{m}$5.1381(1)$^{s}$ & 10\arcm 35.7896(1)\arcs  & 2.39 & 43 & 2041.4$\pm$204  & 12.8$\pm$1.9 & 9.3$\pm$1.9  & 104 & Y\\
J133005+471010$^{\ddagger}$ &30$^{m}$5.1057(1)$^{s}$ & 10\arcm 10.924(1)\arcs & 2.60 & 269  & 4306.7$\pm$431  & 6$\pm$0.1 & - & - & N \\  
J133011+471041&30$^{m}$11.0087(1)$^{s}$ & 10\arcm 41.1393(2)\arcs & 3.27  & 32.7 & 559.7$\pm$63.9  & 6.1$\pm$3.0 & -  & 107 & N  \\ 
J132932+471123&29$^{m}$32.1262(4)$^{s}$ & 11\arcm 23.1175(7)\arcs & 3.52 & 7.1  & 188.1$\pm$51.5 &   8.4$\pm$2.2 & 3.1$\pm$1.5   & - & N \\ 
J133016+471024&30$^{m}$16.0265(1)$^{s}$ & 10\arcm 24.3525(2)\arcs & 4.17  & 20.8 & 455.2$\pm$64.5 &  5.1$\pm$0.6 & -  & - & Y  \\ 
\hline
\multicolumn{10}{l}{$^{\diamond}$at M51a Centre; $^{\ddagger}$SN~2011dh }\\
\end{tabular}
}
\label{tab:VLBI_Sources}
\end{table*}
\end{center}

\section{Multi-wavelength Ancillary Data}

\label{sec:AncillaryData}

To fully understand the nature of the compact sources detected in the wide-field VLBI images, multi-wavelength data were obtained. Here, the data from different facilities that were used to complement the VLBI results are described, including any data processing and analysis that was required.

\subsection{Radio Continuum}

Archival radio data from the VLA and the  Multi-Element Radio Linked Interferometer Network (MERLIN) were obtained for this study. M51 was observed by MERLIN on the 02 July 2005 at 18 cm. The data were calibrated following the MERLIN user guide\footnote{\url{http://www.merlin.ac.uk/user_guide/OnlineMUG-ajh/}}. The entire M51 system ($11^{\prime}.2 \times 6^{\prime}.9$) was imaged in \textsc{aips} using \textsc{imagr}, with Briggs robust parameter 4. Initial imaging of the full MERLIN array revealed a bright $\sim$5~\mjybm ~radio source coincident with the VLBI source J133005+471035 and a weak detection above 300~\mujybm ~(i.e. 5$\sigma$) at the centre of M51. 
By discarding the longest baselines ($>0.8\mathrm{M}\lambda$) and continuum subtracting J133005+471035, deeper images were made revealing a number of radio sources, including diffuse, extended emission around the core, previously detected only with the VLA. MERLIN sources coincident with the VLBI detected sources (\Tab~\ref{tab:VLBI_Sources}) were found via \textsc{blobcat}, using background RMS maps obtained from the source finding algorithm \textsc{aegean} \citep{Hancocketal12}, which were confirmed via visual inspection. The final image gave an rms of 50~\mujybm and a beam size of 0.34\arcs $\times$ 0.17\arcs, PA=-6.8\deg.

Archive VLA-B observations of M51 at 5~GHz (Observation date of 11 December 2003) and VLA-A 1.4~GHz (Observation date of 10 October 2004) were obtained and calibrated as described by \cite{Dumasetal2011}. Both datasets were imaged using the Common Astronomy Software Applications (CASA) package \citep{McMullin+07}, with Briggs robust parameter 2, cell size of 0.3\arcs and image size of 4096 $\times$ 4096 pixels. The imaging procedure also took into consideration w-projection and finally corrected for the primary beam. The final image gave an rms of 12~\mujybm and a beam size of 1.08\arcs $\times$ 1.02\arcs, PA=-87.2\deg for the 1.4~GHz observation and rms of 12~\mujybm and a beam size of 1.72\arcs $\times$ 1.28\arcs, PA=-79.6\deg for the 5~GHz observation. These results will be discussed in the individual source discussions in \Sect~\ref{sec:M51discussion}.

%
%

\subsection{Optical}

Optical images of the M51 system were acquired from the \textit{Hubble Space Telescope (HST)} online archive. A six-pointing  mosaic of the galaxy pair was obtained with the Advanced Camera for Surveys (ACS), Wide Field Channel (WFC) on board the HST in four filters\footnote{HST observing program 10452, P.I. Steven V. W. Beckwith}: B (F435W), V (F555W), I (F814W), and H$\alpha$ + R (F658N) \citep{ACS}. The ACS pipeline-reduced images (with bias subtraction, flat-fielding and drizzling applied) were downloaded from the MAST archive. The final surface brightness of the images are given in units of electrons per second, es$^{-1}$, which can be converted to flux density by the expression

\begin{equation}
F_{\lambda_{\mathrm{eff}}} = C_{\mathrm{es}^{-1}} \cdot 3631 \cdot 10^{-[(Z_{AB}-A_{\lambda})/2.5]},
\label{eq:opticalflux}
\end{equation}

where, $F_{\lambda_{\mathrm{eff}}}$ is the flux density  (in Jansky) at the effective optical wavelength,  $Z_{AB}$ is the zeropoint and defines the absolute physical flux density for a point source of 1~es$^{-1}$ for a particular ACS-WFC filter \citep{Siriannietal05}, and $A_{\lambda}$ is the Galactic foreground colour magnitude extinction towards M51. The values of $Z_{AB}$ and $A_{\lambda}$ for the different HST filters are listed in \Tab~\ref{tab:HSTfilter}. The parameter $C_{\mathrm{es}^{-1}}$ is the net count rate for an infinite aperture. We used standard Heasoft tasks to carry out aperture photometry on the optical sources with an aperture corrective given by \citet{Siriannietal05}. Finally, using the relations given by \citet{Siriannietal05} we converted between ACS-WFC $F_{\lambda_{eff}}$ and the Johnson-Cousins \textit{UBVRI} photometric systems.

\begin{center}
\begin{table}\footnotesize
\caption{\textsc{Parameters of the HST ACS Filters}}
\label{tab:HSTfilter}
\hfill{}
\begin{tabular}{ccccc}
\hline\hline
\noalign{\smallskip}
 &  & $\lambda_{\mathrm{eff}}$ & $Z_{AB}^{\ast}$ & $A_{\lambda}^{\dagger}$ \\ 
Filter & Colour & (\angstrom) &(mag) &  (mag) \\ 
 \hline
F435W & B & 4450 & 25.67 & 0.125  \\ 
F555W & V & 5510 & 25.72 & 0.091 \\ 
F814W & I & 8060 & 25.94 & 0.053 \\ 
F658N & H$\alpha$ & 6563 & 22.77 & 0.08 \\ 
\hline
\multicolumn{5}{l}{$^{\ast}$\citealt{Siriannietal05}; $^{\dagger}$ Obtained from the }\\
\multicolumn{5}{l}{ NASA/IPAC  Extragalactic Database}
\end{tabular}
\hfill{}
\end{table}
\end{center}

\subsection{X-rays}
\label{sec:xrays}

M51 has been observed by \textit{Chandra/ACIS-S} 12 times (considering
only observations longer than 2ks) between 2000 and 2012 (\Tab~\ref{tab:Chnadra_data}).
Aim points and roll orientations differed between observations,
but the nucleus of M51a was always located in the field of the back-illuminated
S3 chip. We downloaded the data from the public {\it Chandra} archive.
We reprocessed each observation using tools available in {\small {CIAO}}
Version 4.5 \citep{Fruscioneetal06}, and CalDB version 4.5.3, to obtain new level-2 event files.
We then reprojected the event files to a common tangent point using
the {\small {CIAO}} script {\it reproject$\_$all}. The total exposure time
in the merged file is 855 ks for most of the D25 region of M51a.
We used the merged event file for imaging and source detection purposes.
For spectral analysis of the brightest sources, we generated separate
PHA, RMF, and ARF files for each observation with the {\small {CIAO}}
task {\it specextract}, and created combined spectra by averaging
and merging the spectra from the individual observations.
We then fitted the spectra with standard models in XSPEC  Version 12.6
\citep{Arnaud96}. For sources with sufficiently high S/N,
we binned the spectra to $\ge 15$ counts per channel and used $\chi^{2}$
statistics. For sources with $\la 100$ counts, we used Cash statistics
\citep{Cash79}.

\Tab~\ref{tab:Chnadra_detections} lists the X-ray detections for the sources in \Tab~\ref{tab:VLBI_Sources} and other sources discussed in this paper: SN~1945A; SN~1994I; SN~2005cs; and Component N.
%

\begin{center}
\begin{table}\footnotesize
\caption{\textsc{Log of the chandra observations}}
\label{tab:Chnadra_data}
\hfill{}
\begin{tabular}{ccc}
\hline\hline
\noalign{\smallskip}

ObsID & Obs.~Date & Exp.~Time    \\
 & (dd/mm/yy) & (ks)   \\ 
(1) & (2) & (3)   \\ 
\hline
354  & 20/06/00  & 14.86  \\ 
1622  & 23/06/01  & 26.81  \\ 
3932 & 07/08/03  & 47.97  \\ 
12562 & 12/06/11  & 9.63  \\ 
12668 & 03/07/11  & 9.99   \\ 
13812 & 09/09/12  & 179.2   \\ 
13813 & 12/09/12  & 157.46  \\ 
15496 & 19/09/12  & 40.97  \\ 
13814 & 20/09/12  & 189.85  \\ 
13815 & 23/09/12  & 67.18  \\ 
13816 & 26/09/12  & 73.1  \\ 
15553 & 10/10/12  & 37.57  \\ 
\hline
\end{tabular}
\hfill{}
\end{table}
\end{center}

\begin{center}
\begin{table}\footnotesize
\caption{\textsc{Radio/X-ray associations}}
\label{tab:Chnadra_detections}
\hfill{}
\begin{tabular}{cccc}
\hline\hline
\noalign{\smallskip}
&\multicolumn{ 2}{c}{Position (J2000.0)} &    \\ 
\cline{2-3}
 & RA & Dec &  \\ 
VLBI Sources &13$^{h}$ & 47\deg  & X-rays?  \\ 
\hline
J132952+471142$^{\diamond}$  & 29$^{m}$52$^{s}$.7102 & 11\arcm 42\arcs.7465  &  YES \\ 
J133005+471035 & 30$^{m}$05$^{s}$.1381  & 10\arcm 35\arcs.7896  &  YES$^{\star}$ \\ 
J133005+471010$^{\ddagger}$ & 30$^{m}$05$^{s}$.1057  & 10\arcm 10\arcs.924    &  YES \\ 
J133011+471041 & 30$^{m}$11$^{s}$.0087 & 10\arcm 41\arcs.1393  &  YES \\ 
J132932+471123 & 29$^{m}$32$^{s}$.1262 & 11\arcm 23\arcs.1175 &  YES$^{\star}$ \\ 
J133016+471024 & 30$^{m}$16$^{s}$.0265 & 10\arcm 24\arcs.3525 &  YES \\ 
\hline
Other Sources & RA & Dec & X-rays? \\ 
\hline
SN~1945A &  29$^{m}$ 58$^{s}$.95 & 15\arcm 53\arcs.8 &  NO \\ 
SN~1994I &  29$^{m}$ 54$^{s}$.072 & 11\arcm 30\arcs.5  &  YES \\ 
SN~2005cs & 29$^{m}$ 52$^{s}$.850 &  10\arcm 36\arcs.3  &  unclear \\ 
Component N & 29$^{m}$ 51$^{s}$.59 &  12\arcm 07\arcs.7  &  YES$^{\star}$ \\ 
\hline
\multicolumn{4}{l}{$^{\diamond}$Nuclear source; $^{\ddagger}$SN~2011dh; $^{\star}$ very faint}\\
\end{tabular}
\hfill{}
\end{table}
\end{center}

\section{Discussion}

\label{sec:M51discussion}

\subsection{Fraction of Radio Sources Detected}
\label{sec:frac}
Of the 107 radio sources detected in M51 by the \citet{Maddoxetal07} VLA survey (beam size: 1.50\arcs$\times$1.21\arcs, 21~cm;  1.47\arcs$\times$1.13\arcs, 6~cm. Sensitivity: 22.5\mujybm, 21~cm; 11.7\mujybm, 6~cm), we report three confirmed detections. Similarly, of the 15 sources in the same region detected by the VLA FIRST survey \citep{FIRST95}, three were detected\footnote{Two sources were detected by both surveys - the core of M51 and source J133005+471035}. For a more realistic estimate of the proportion of detected sources we restrict ourselves to the catalogued sources that are bright enough to be detected above the 6.7$\sigma$ threshold of the EVN observation. The catalogued flux densities were adjusted to 18~cm by using the spectral indices provided by the catalogues, and the sensitivity of the EVN phase centres were corrected for primary beam effects. This leaves 70 sources from the \citet{Maddoxetal07} VLA survey and 15 from the FIRST survey. Assuming no catalogued sources are resolved, the fraction of VLBI sources detected compared to the \citet{Maddoxetal07} and  VLA FIRST surveys are 3/70 ($^{+0.12}_{-0.02}$) and 3/15 ($^{+0.46}_{-0.07}$) \footnote{Population uncertainties were derived using the Bayesian beta distribution quantile technique \citep{Cameron2011}.}, respectively. 

Considering sources from the \citet{Maddoxetal07} survey not associated with HII regions, the detection rate is 3/32 ($^{+0.24}_{-0.03}$), while the VLBI detection rate for sources with peak flux density $>$ 1~mJy in the \citet{Maddoxetal07} survey is 2/5 ($^{+0.77}_{-0.12}$). The fraction of sources detected from the FIRST survey (3/15) and \citet{Maddoxetal07} sources with flux density $>$ 1~mJy (2/5), closely agrees with the detection rates of previous wide-field VLBI surveys (e.g. \citealt{Middelbergetal11, Morganetal2013, Middelbergetal2013}). However, the fraction of  \citet{Maddoxetal07} sources not associated with HII regions detected with VLBI (3/32) is significantly lower than the previous wide-field VLBI  studies.

At 1.6~GHz, our observations would be less affected by free-free absorption compared to \citet{Maddoxetal07} 1.4~GHz observations (e.g \citealt{Tingay04}). It is likely that the majority of the radio sources in M51 detected by \citet{Maddoxetal07} are fully resolved at our VLBI resolution, suggesting perhaps either lack of compact radio emission from sources associated with stellar clusters or the expansion of supernova remnants (SNRs) beyond the maximum size detectable with our observations. 

\citet{Maddoxetal07} estimated that the typical sizes for SNRs in M51 range between 0.3$^{\prime\prime}$ and 0.5$^{\prime\prime}$, (i.e. 12.2 - 20.4 pc). Considering the maximum source size resolvable with our observations is 0.14$^{\prime\prime}$  ($\sim$6 pc; see \Sect~\ref{sec:obs}), the non-detections are not surprising.  Assuming an expansion velocity of 21,000 
km~s$^{-1}$ for core-collapse supernovae in M51 \citep{Weileretal11, Bietenholzetal2012} gives an age of 570 - 950 yr for the largest and 280 years for the smallest shell sources.

\subsection{The M51a Nuclear Region}

\subsubsection{A Faint Seyfert Radio Nucleus}

The VLBI observation resolves the radio source at the nucleus of M51a into multiple components (source J132952+471142 in \Fig~\ref{fig:VLBI}a). The main feature is a compact source (labelled C1) marginally resolved in the high resolution images (black contours), and unresolved in the 10~M$\lambda$ image. Located to the north of C1 are two weaker sources (labelled C2a and C2b), which are blended into a single, unresolved elongated structure that appears to be connected to C1 in the 10~M$\lambda$ image.  Components C2a and C2b are unresolved sources with total flux densities of 51.5 $\pm$ 9.1 \muJy ~and 54.6 $\pm$ 10.2 \muJy, respectively. The separation of C2a and  C2b from C1 are $\sim$ 41.5 mas ($\sim$ 1.69 pc) and  $\sim$ 49.4 mas ($\sim$ 2.01 pc), respectively.

VLBI observations of a distance-limited sample ($<$22~Mpc) of Seyfert galaxies, at 18 and 6~cm, have revealed similar structures to M51a within their nuclei \citep{GP09, Bontempietal12, PG13}. Here, the detected VLBI flux density is dominated by a high brightness temperature (log$_{10}$T$_{B}>7.5$) parsec-scale source with a flat/intermediate spectrum ($0.3\leq\alpha\leq0.6$), accompanied by fainter, steep spectrum extended components. The flat/intermediate spectrum, high T$_{B}$ sources are taken as  evidence for non-thermal processes driven by jet-producing central engines typical of radio-quiet and radio loud quasars and AGNs \citep{BB98,Doietal2013}. 

Using the brightness temperature relation, T$_{B}$ =$ \dfrac{S(\nu)}{2k\theta_{\mathrm{maj}}\theta_{\mathrm{min}}}\, \left(  \dfrac{c}{\nu}\right) ^{2}$, where: $S(\nu)$ is the flux density at frequency $\nu$; $k$ - the Boltzmann constant; $\theta_{\mathrm{maj}}\,\rm and\, \theta_{\mathrm{min}}$ are the major and minor deconvolved sizes in radians; and c is the speed of light, a value of log$_{10}$T$_{B}\,\sim\,6.6$ is obtained for C1. Both T$_{B}$ and luminosity (10$^{18}$ W Hz$^{-1}$) of C1 are within the typical range of values for low luminosity AGNs within Seyfert galaxies \citep{PG13}. Therefore, components C2a and C2b are, most probably, hotspots associated with a jet or thermal outflow.
 
\subsubsection{Previous VLBI observations}
\label{sec:M51archiveVLBI}
\citet{Bontempietal12} observed M51a with the EVN at both 18 and 6~cm and were unable to detect any emission above 3$\sigma$ (75 and 160 \mujybm ~at 18 and 6~cm, respectively). However, they searched within a 1\arcs radius centred on the VLA FIRST position RA = 13$^{\mathrm{h}}\,29^{\mathrm{m}}\,52.804^{\mathrm{s}}$, Dec = +47\deg 11\arcm 40.065\arcs,
which would mean that the VLBI position in \Tab~\ref{tab:VLBI_Sources} was beyond their search field. 
The uncalibrated datasets for both observations (Experiment ID: EG037D and EG037C. PI: Giroletti) were obtained from the EVN data archive. Using the calibration tables generated by the EVN pipeline, both datasets were imaged as described in \Sect~5.3.1. 

The core-jet structure was detected only in the 18~cm data at the expected position with a similar size, flux density and morphology as seen in the wide-field observation presented in \Sect~\ref{sec:M51Fluxes}. The non-detection at 5~GHz is attributed to the poor sensitivity of the observation ($\sigma$=50 \mujybm), caused by the loss of Ef for 90$\%$ of the time due to high winds. Assuming a flat to intermediate spectrum ($-0.5\leq\alpha\leq0.5$), the expected flux density of J132952+471142 at 6~cm would be less than 150 \mujybm.

\begin{figure}
\centering
\includegraphics[trim=0cm 1.7cm 0cm 0cm,clip=true,scale=0.40, angle=-90]{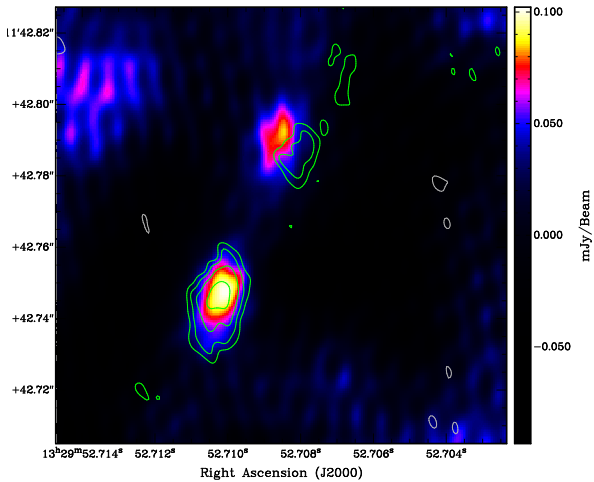} 
\caption{VLBI images of the Seyfert nucleus of M51a for the 2008 epoch  (colour) with the \citet{Bontempietal12} dataset and the 2011 epoch (contours). Both epochs were imaged with a beam size of 18.5 mas $\times$ 11.0 mas with a P.A. = -3.71\deg. The RMS noise, $\sigma$ in both images are 19.2~\mujybm ~and 8.9~\mujybm ~for the 2008 and 2011 images respectively. While the peak fluxes are 102.3 $\pm$ 21.0~\mujybm ~(2008) and 84.7 $\pm$ 14.7 \mujybm ~(2011). The first contours of the 2011 image are at $\pm3\sigma$, and increase in multiples of $\sqrt{2}$. The offset in position of the low surface brightness structures  between epochs is 7.53 mas. A third EVN observation has been requested to investigate the nature of this offset.}
\label{fig:M51jet}
\end{figure}

The images from both epochs (2008 and 2011) at 18~cm are compared in \Fig~\ref{fig:M51jet}, where the contours of the 2011 epoch are overlayed on the 2008 epoch (colour image). A restoring beam of size 18.5 mas $\times$ 11.0 mas with a P.A. = -3.71\deg was used for the 2011 epoch so as to match the 2008 epoch. While the compact radio core is found at the same position in both epochs, the lower brightness compact source associated with  a jet/thermal outflow shows an apparent lateral shift in position of 7.53 mas between epochs. The positional measurementsfor both epochs were obtained using the same phase reference source, with the same referenced position.

With a time difference of 3.7 years, and assuming that the difference in position is genuine, a velocity of 0.27c is implied. Furthermore, the peak brightness of the jet/hotspot has decreased between epochs (80.2 \mujybm, 2008; 44.2 \mujybm, 2011), which may be associated with the apparent motion. Molecular aperture synthesis HCN observations of M51a by \citet{Kohnoetal96} revealed a twisted velocity field resulting from the existence of a dense rotating molecular disk that is aligned to the nuclear jet of M51a, and the circular rotation of the galactic disk plane. However, the maximum velocity of this field (450-500~km~s$^{-1}$) is too low to result in the apparent motion of the pc-scale jet. Furthermore, the direction of the apparent motion suggests that the jet may be precessing. While jet precession is not uncommon in radio bright AGNs (e.g \citealt{CM93,Appletal96,LZ05}), it is yet to be observed for low luminosity Seyfert galaxies. In addition, outward proper motions have been observed (at the 3$\sigma$ level) in the jets of the Seyfert galaxy NGC~5506 on pc scales \citep{Middelbergetal04}. The apparent motion could be the result of a combined effects of forward bulk motions of the jet interacting with the rotation observed by \citet{Kohnoetal96}.

It is, however, possible that the position shift can be the result of systematic factors (e.g. poor calibration), since the hotspots displaying the apparent motions are $\sim$4$\sigma$ detections. This is unlikely since the positions of the phase calibrator and C1 between both epochs were found to agree to within less than 1 mas. Whether this apparent motion is indeed intrinsic to M51a is difficult to ascertain with only two epochs. Therefore, a request has been made to observe the nuclear region of M51a with a third higher sensitivity observation with the EVN at 18~cm to determine whether this is indeed intrinsic to the parsec-scale jet of M51a, the results of which will be discussed in a later paper.

\subsubsection{A Multi-wavelength Perspective}

\begin{figure*}
\centering
\includegraphics[width=\textwidth]{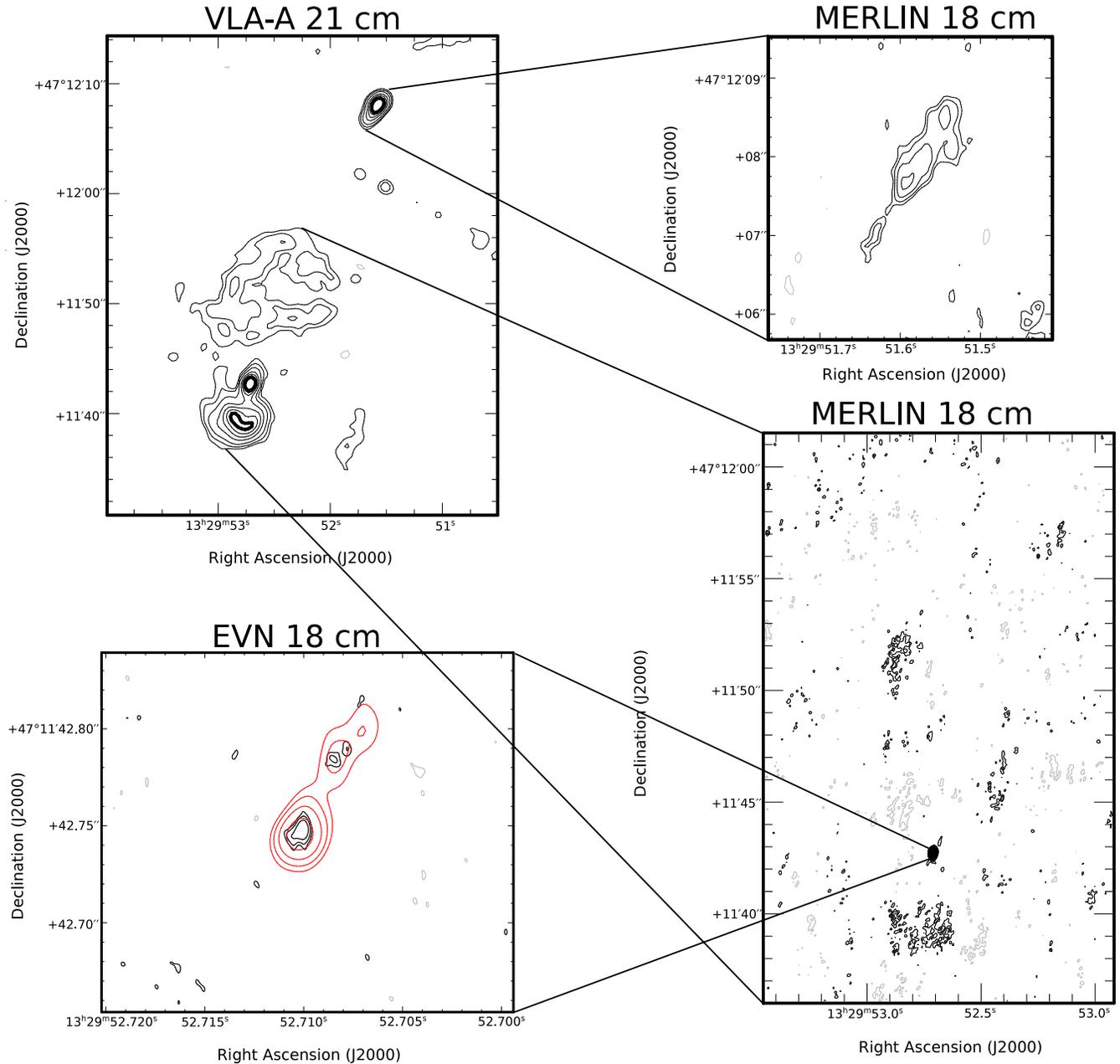} 
\caption{Radio images of the nuclear region, XNC, northern bubble and a bright source located $\sim$ 30\arcs ($\sim$1.22 kpc) from the core. \textit{VLA-A}: The first contours are at $\pm15\%$ the peak flux (11.49\mjybm, and increase in increments of $10\%$ the peak flux;  \textit{MERLIN}: The first contours are at $\pm3\sigma$, and increases in multiples of $\sqrt{2}$, where $\sigma$ = 50$\mu$Jy beam$^{-1}$; \textit{EVN}: see \fig~\ref{fig:VLBI}a. The negative contours are plotted in grey.}
\label{fig:M51montage}
\end{figure*}

\Fig~\ref{fig:M51montage} shows the complex nuclear region of M51a with increasing resolution. The VLA-A image at 20 cm (resolution $\sim$1.08\arcs) and the higher resolution MERLIN 18~cm images (resolution $\sim$0.34\arcs) reveal an area of ring-like emission to the north of the core (hereafter, the northern bubble) and a dense area of emission to the south, referred to as an extranuclear cloud (XNC), that may have resulted from interactions of a nuclear jet with the surrounding ISM \citep{CvH92,TW01, Maddoxetal07}. \citet{Maddoxetal07} measured peak flux densities for the nucleus at both 20~cm (2.08~\mjybm ) and 6~cm (1.14~\mjybm), giving a moderately flat spectrum ($\alpha$ = $-$0.49). The MERLIN image shows the more compact regions where the ejecta are strongly interacting with the surrounding medium. Also shown in the image is a slightly extended source located $\sim$30\arcs ($\sim$1.22 kpc) from the nuclear radio source, which is discussed later in \Sect~\ref{sec:relic}.

\medskip

As noted by previous studies \citep{TW01, Maddoxetal07}, the radio continuum emission of the nuclear region, including the XNC and the 
northern bubble is very similar to the \textit{Chandra} X-ray morphology. \citet{TW01} found that the X-ray spectrum mainly consist of a hard component that  dominates the 3-8 keV band corresponding to an obscuration of the nuclear source by a column density in excess of $10^{24}$ cm$^{-2}$. With our deeper combined spectrum (855 ks), we confirm that 
the emission consists of a soft thermal-plasma component 
plus a hard component probably due to reflection or scattering, 
with a strong Fe line. 
The average observed flux in the 0.3--10 keV band 
is $f_{0.3-10} = (1.4 \pm 0.1) \times 10^{-13}$ erg cm$^{-2}$ s$^{-1}$, 
of which $\approx 3.5 \times 10^{-14}$ erg cm$^{-2}$ s$^{-1}$ 
is in the thermal-plasma component, which may come from a compact 
nuclear starburst or from gas shock-ionized by nuclear outflows. 
The unabsorbed luminosity of the thermal-plasma emission 
is $L_{0.3-10} \approx 3.3 \times 10^{38}$ erg s$^{-1}$. 
For the hard component, nominally we obtain 
$L_{0.3-10} = (9 \pm 1) \times 10^{38}$ erg s$^{-1}$; 
however, this is only a minimal fraction of the true X-ray luminosity 
of the nuclear black hole, whose direct emission is completely obscured 
in the {\it Chandra} energy band. Using {\it BeppoSAX}, 
\citet{Fukazawaetal01} detected the direct hard X-ray emission 
at energies $\ga 10$ keV, and estimated a column density 
$N_{\rm H} \approx 6 \times 10^{24}$ cm$^{-2}$ and an intrinsic nuclear luminosity 
$L_{2-10} \approx 10^{41}$ erg~s$^{-1}$. 
More detailed analysis of the nuclear X-ray spectrum is beyond the scope 
of this work.

The X-ray emission of both the XNC and the northern bubble is dominated by optically-thin thermal plasma at a characteristic temperature $kT \approx 0.6$~keV, typical of collisionally-ionized gas \citep{TW01}. High-resolution 6~cm \citep{CvH92} and 3~cm \citep{Bradleyetal2004} VLA observations of the nucleus reveals a narrow jet-like feature emanating from the nucleus and terminating in the XNC. [O \textsc{iii}] line measurements of narrow line regions within the nuclear region, and along the direction of the narrow jet-like feature, find evidence for shock heating of the gas \citep{Bradleyetal2004} at the location of the XNC, indicating continuous fuelling of the XNC by the radio jet. However, no evidence has been found for shock ionisation in narrow line regions within the direction of the northern bubble \citep{Bradleyetal2004}, suggesting the northern bubble may therefore be from a previous ejection cycle of the Seyfert nucleus of M51a \citep{CvH92,Maddoxetal07}. Interestingly, jet hotspots were detected with VLBI only in the direction of the northern bubble (components C2a and C2b). However, the AGN core appears slightly elongated in the VLBI image, in the direction of the southern jet, and may be the launching point for the jet.

\subsubsection{Component N: a possible fossil radio hotspot}
\label{sec:relic}

The VLA and MERLIN images reveal a bright object (2229$\,\pm\,$21 \mujybm, VLA 20~cm; 946$\,\pm\,$11 \mujybm, VLA 6~cm; 448.3$\,\pm\,$54 \mujybm, MERLIN 18~cm;) located to the north of the core of M51a that was not detected in the EVN observations. The MERLIN position (RA = 13$^{\mathrm{h}}$ 29$^{\mathrm{m}}$ 51.593$^{\mathrm{s}}$, Dec = 47\deg 12\arcm 07.66\arcs) places Component N $\sim$25.85\arcs from the nearest phase centre. The unaveraged visibilities for the phase centre closest, was shifted to the position of Component N. The dataset was then averaged in time and frequency, such that the combined smearing at a radius of 2\arcs from the MERLIN position was limited to 5$\%$. The resulting dataset was imaged also with uv-tapering of 10 M$\lambda$, with natural weighting, which resulted in no detection above 5$\sigma$. Given that the primary beam sensitivity at component N is is 5-10$\%$ the peak sensitivity, we estimate an upper limit of the EVN peak flux density of 82.5~\mujybm. 

The object, hereafter component N, has a steep spectrum ($\alpha^{20cm}_{6cm}$ = -0.7), dominated by synchrotron radio emission, thus suggesting it is possibly a jet hotspot associated with the Seyfert nucleus of M51a, a SNR or a background AGN. The MERLIN 18~cm image in \Fig~\ref{fig:M51montage} reveals a complicated structure, elongated in the direction of M51a nucleus, atypical of SNRs. Moreover, the lack of detection of high brightness temperature emission with the EVN suggests it may be a background AGN whose core has either been scatter broadened or a background galaxy without a compact radio core.

\begin{figure*}\scriptsize
\centering
\subfloat{\includegraphics[scale=0.35]{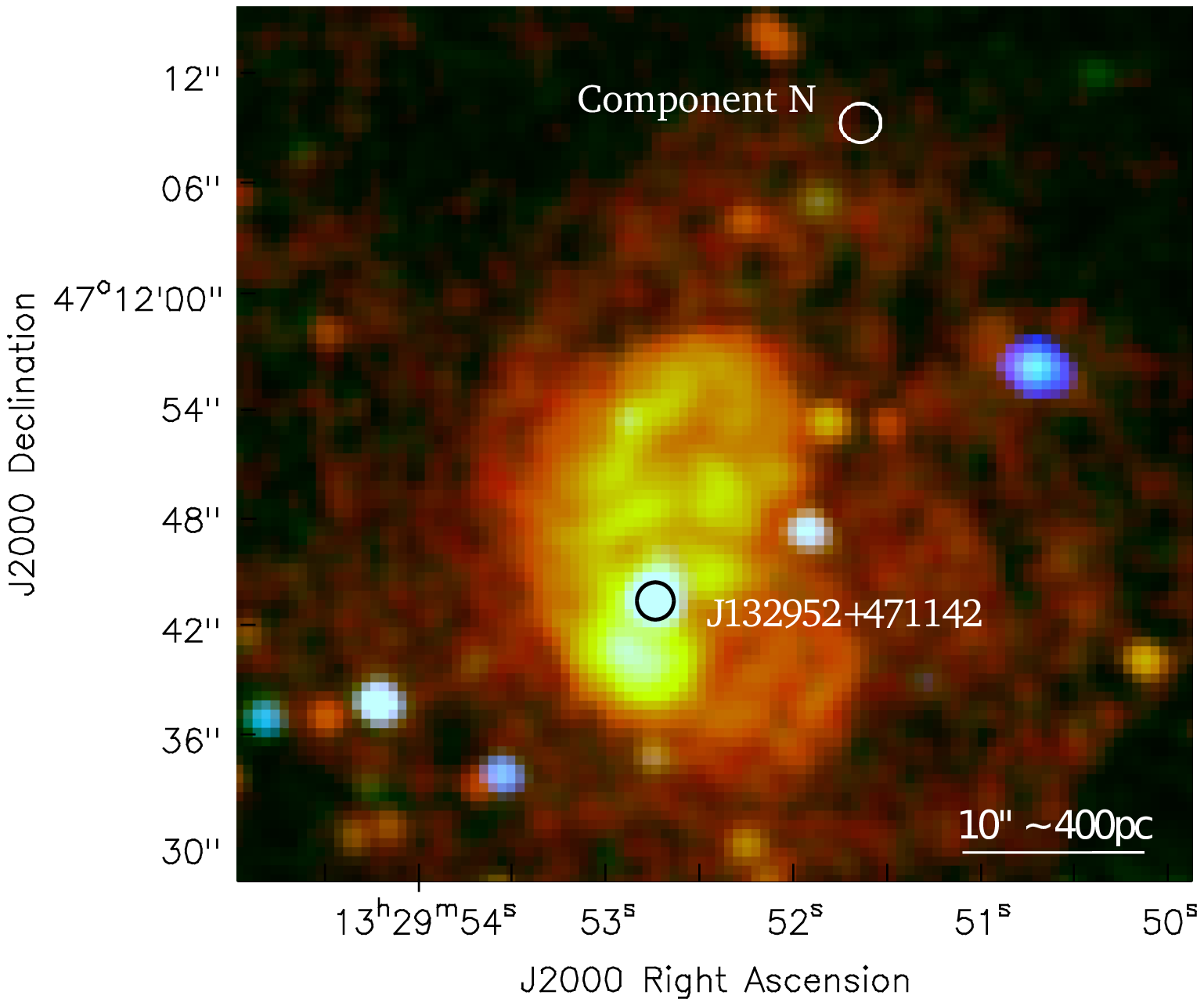}}\hspace{-2em}%
\subfloat{\includegraphics[scale=0.35]{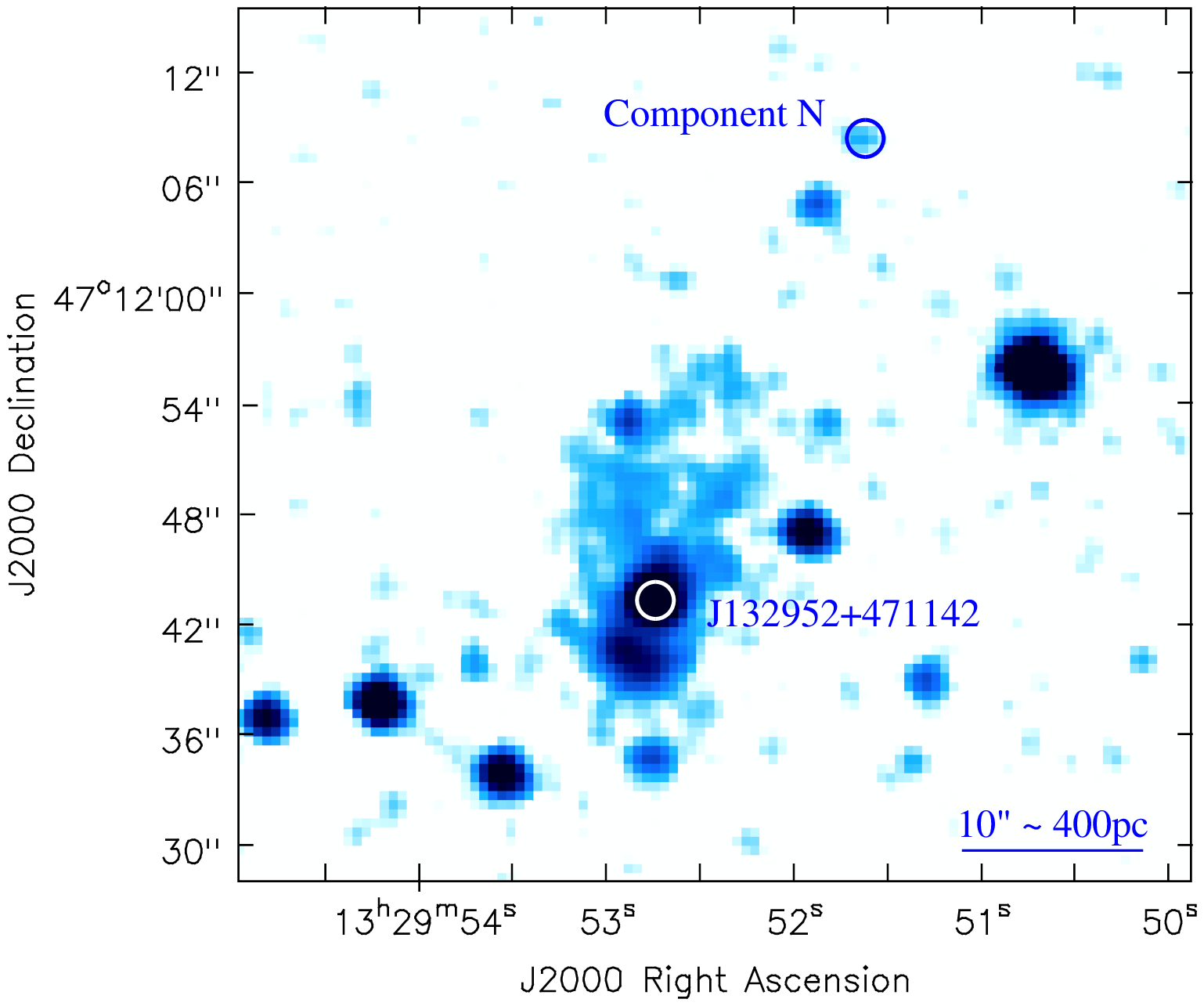}}\\[-2ex]
\subfloat{\includegraphics[scale=0.35]{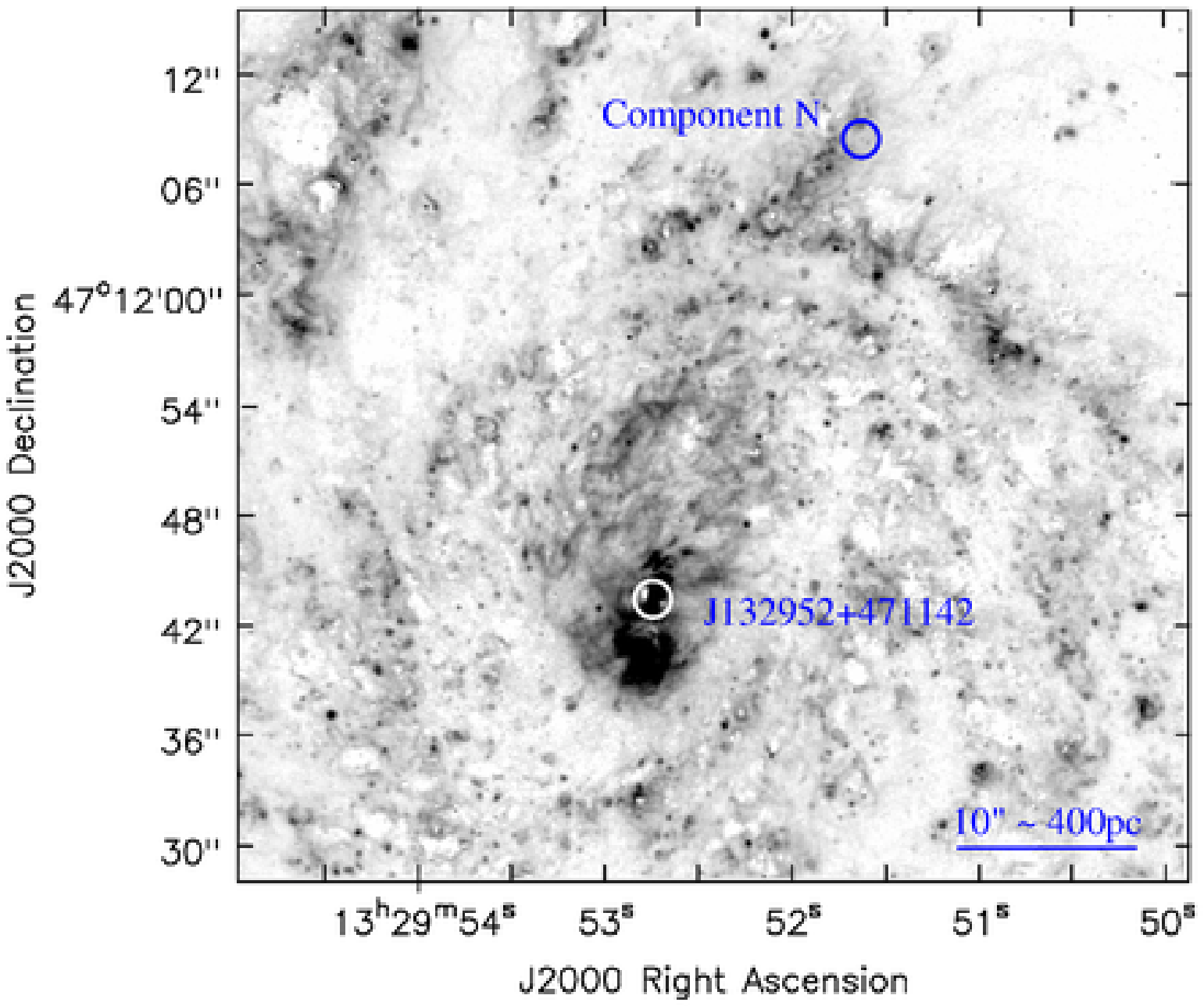}}\hspace{-2em}%
\subfloat{\includegraphics[scale=0.35]{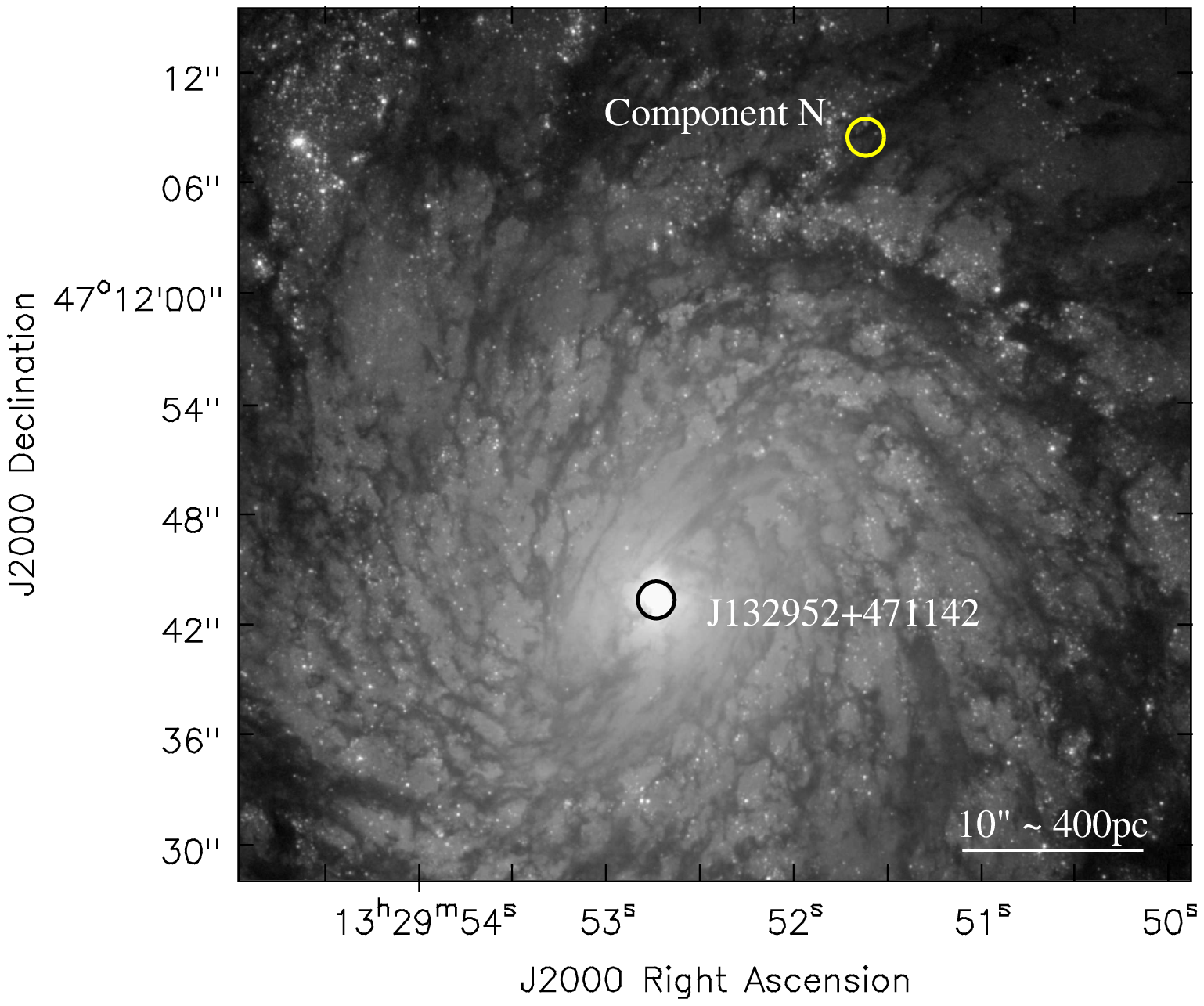}}
\caption{{\bf Top left}: {\it Chandra}/ACIS-S X-ray image of the nuclear 
region of M51a. Colours are: red = 0.3--1 keV, green = 1--2 keV; 
blue = 2--7 keV. The location of the radio nucleus (coincident 
with the X-ray nuclear source) and of MERLIN's Component N 
are circled. The radius of the circles is 1\arcs; this is chosen 
for display purposes only, and is much larger than the positional uncertainty.
Note the similar morphology of the diffuse soft X-ray emission with the 
diffuse radio emission (Figure 11).
{\bf Top right}: {\it Chandra}/ACIS-S X-ray image of the same field, 
in the hard band only (2--7 keV). The image has been smoothed 
with a 2-pixel Gaussian. Nucleus and Component N are circled. 
{\bf Bottom left}: {\it HST}/ACS continuum-subtracted image in the F658N 
filter (H$\alpha$) of the same field, with nucleus and Component N circled. 
{\bf Bottom right}: {\it HST}/ACS image in the F555W 
filter (V band), with nucleus and Component N circled.}
\label{fig:X_ray_Component_N}
\end{figure*}

There is a faint, point-like X-ray source (\fig~\ref{fig:X_ray_Component_N}, 
top right panel) at the position of Component N in the {\it Chandra} data. 
The source is significantly detected in the hard band (2--7 keV). 
The detection is only at the 3$\sigma$ level; however, 
the exact positional coincidence with the radio source gives us confidence 
that it is a real X-ray source.
At softer energies, such a faint source is swamped by the diffuse 
thermal-plasma emission in that region (\fig~\ref{fig:X_ray_Component_N}, 
top left panel); but we cannot rule out the possibility that the point-like 
X-ray source is heavily absorbed below 2 keV, perhaps by one of 
the numerous dust lanes in that region. {\it HST} continuum images 
do in fact suggest the presence of a dust feature at the location 
of Component N  (\fig~\ref{fig:X_ray_Component_N}, bottom right panel).
Based on the {\it Chandra}/ACIS 2--7 keV 
count rate of $(2.0 \pm 0.6) \times 10^{-5}$ ct s$^{-1}$, and assuming 
a power-law photon index $\Gamma = 1.7$ and line-of-sight absorption, 
we estimate an observed 0.3--10 keV flux 
$f_{0.3-10} = 1.1^{+0.2}_{-0.3} \times 10^{-15}$ erg cm$^{-2}$ s$^{-1}$
and an emitted luminosity 
$L_{0.3-10} = 1.0^{+0.2}_{-0.3} \times 10^{37}$ erg s$^{-1}$ if located in M\,51.

We inspected the {\it HST}/ACS-WFC images but did not find 
any obvious counterpart of Component N in the continuum 
(\fig~\ref{fig:X_ray_Component_N}, bottom right panel); 
a few nearby point-like sources are consistent with small, unresolved 
(FWHM $\la$ 0\arcs1 $\approx 40$ pc) groups of young massive stars, 
abundantly scattered in the inner disk of M51a. An elongated stream of
H$\alpha$ emission near Component N is oriented towards the nucleus  
(\fig~\ref{fig:X_ray_Component_N}, bottom left panel), roughly 
in the same direction as the putative jet, but we cannot determine 
at this stage if there is a physical relationship. Further 
analysis of the circumnuclear H$\alpha$ and continuum emission is beyond 
the scope of this paper.

Possible interpretations for the Component N radio and X-ray sources
are synchrotron emission from a hot spot energized by a nuclear jet, or 
a background radio galaxy seen through the disk of M51a. 
If it is a background AGN, it is a remarkable coincidence that its radio 
structure is elongated towards the core of M51a, but that is not a priori 
impossible (a similar situation is well known in the spiral galaxy M\,83, 
with a bright background FRII radio galaxy projected near the nucleus, which 
might at first sight be interpreted as a jet feature from M\,83 itself \citep{Maddoxetal2006, Longetal14}.

To determine the probability of a chance alignment of a background AGN with the M51a nucleus we used the method of estimating quasar alignment by \citet{EG85}. In this method, \citet{EG85} investigated the expected number of random alignments, $\mu$, in a field of area $A$, with a surface density of sources, $n$ for random clustered and unclustered fields and found it to be approximately,

\begin{equation}
\mu=\dfrac{2\pi}{3} p_{\mathrm{max}} d^{3}_{\mathrm{max}} A n^{3},
\label{eq:mudensity}
\end{equation}

where $d_{\mathrm{max}}$ is the distance between the two furthest sources, $p_{\mathrm{max}}$ is the distance between the outer two sources, such that the product $d_{\mathrm{max}}\cdot 2p_{\mathrm{max}}$ forms a rectangle enclosing all points of interest. The number density of quasars from the FIRST survey is $\sim$90 quasars per square degree \citep{Whiteetal1997} and the total area surveyed in this study is 11.2\arcm$\times$6.9\arcm. Thus, using \eqn~\ref{eq:mudensity}, $\mu \ll 1$ for the inner region of M51a i.e. $d_{\mathrm{max}}$ = 30\arcs and $p_{\mathrm{max}}$ = 6\arcs. There is, therefore, a very low probability of a chance alignment within the central region of M51a, suggesting that component N and the nucleus of M51a are most probably physically related. Thus, component N is most likely a jet hotspot associated with the Seyfert nucleus of M51a.

However, with no evidence of a continuous jet in the northern region of M51a, and the lack of a high brightness temperature component, Component N may be a fossil radio lobe/hotspot associated with past AGN activity from the core of M51a. Fossil radio lobes have been detected in a few radio galaxies such as 3C~388 \citep{JP01,Gentile2007} and IC~2476 \citep{Cordey1987}, with a few ``dying galaxies" with evidence for fossil radio lobes (e.g. \citealt{Parmaetal2007, Murgiaetal12}). Radio lobes or hotspots are powered by energy from an AGN via jets of plasma. When the AGN undergoes a fading or dying stage, the jets and eventually the hotspots will expand adiabatically into the surrounding medium, and disappear due to the loss of the continuous production of plasma \citep{Murgiaetal12}. It is possible that this eventually leads to the loss of the high brightness temperature component, leaving behind a diffuse shell. 

If this is the case then the remaining synchrotron radiation in component N would be from electrons older than those within the core (e.g. \citealt{Carillietal1991}). Synchrotron ageing in radio galaxies is generally associated with a spectral ``break" that shifts to lower frequencies with older plasma \citep{Carillietal1991}. To investigate this possibility, the inner region on M51a (including Component N) will be observed with the Low Frequency Array (LOFAR) including all the international stations in VLBI mode at 110 and 170 MHz.

\subsection{Supernovae}

At the time of our observations, M51a and M51b hosted four optically observed supernovae (SNe): the Type Ia SN~1945A \citep{KS1971}; the Type Ib/c SN~1994I \citep{Puckett1994}; the Type Ib/c SN~2005cs \citep{Muendlein2005}; and the Type IIb SN~2011dh \citep{Griga2011}. Of the four SNe, radio emission and X-rays have been detected in SN~1994I \citep[Radio]{Weileretal11}, \citep[X-Rays]{Immler+02,Immler+98} and SN~2011dh \citep[Radio]{Krauss2012,Bietenholzetal2012,Horeshetal13}, \citep[X-Rays]{Soderbergetal12, CS12, Maeda+14}. 

SN~2005cs is located within a region with a number of faint sources near the detection limit. There is however, soft X-ray emission inside a 1\arcs error circle, from at least two sources with $\approx$20--30 counts, but it is unclear whether either of them is the true X-ray counterpart of SN~2005cs. In any case, we can say that the X-ray luminosity $L_{0.3-10} \la 2 \times 10^{36}$ erg~s$^{-1}$, which is two orders of magnitude lower than the upper limits on the X-ray luminosity estimated by \cite{Brown+07}. We will now discuss the radio and x-ray properties of, SN~1994I and SN~2011dh.

\subsubsection{SN~1994I}

\textbf{Radio} Almost a decade after the explosion, \citet{Maddoxetal07} detected radio emission from SN~1994I. However, no radio emission above the detection threshold of 50.0~\mujybm ~ was found in the VLBI images at the position of SN~1994I. Given that our observation was almost a decade from the \citet{Maddoxetal07} observation, it is possible that SN~1994I has faded below our sensitivity and/or expanded beyond the EVN's maximum resolvable size. Adjusting the 1.4~GHz flux density from \citet{Maddoxetal07} with a spectral index of -1.04 \citep{Maddoxetal07}, the flux density of SN~1994I at 1.65~GHz at the time of the \citet{Maddoxetal07} observation would have been 134.9\mujybm. The supernova light curve model of \citet{Weileretal02} indicates that the flux density, $S$ decreases with time following a power-law, $S\,\propto\,t_{age}^{\beta}$, where $t_{age}$ is supernova age and $\beta$=-1.42  for SN~1994I \citep{Weileretal11}. At the time of our observation (age of 6428 days) a 1.65~GHz peak flux density of 59.5~\mujybm ~is estimated, which is just above our detection threshold. Furthermore, the expansion of SN~1994I has been well studied and follows the form 1.29($t_{age}$/1~day)~$\mu$as \citep{Weileretal11}, which estimates a size of 8.3~mas at the time of our observation. The estimates of size and flux density are just within the limits of our EVN observation; as these estimates are model dependent, it is possible that SN~1994I was just below the detection threshold. It is also possible that the power-law decrease of the flux density steepens at later times, reducing the flux density below our detection threshold. This has been observed for SN1957D, albeit more than 40 years after its explosion \citep{Long+12}.

\textbf{X-Rays.} SN~1994I was detected in the first three Chandra observations 
(\tab~\ref{tab:Chnadra_data}) with declining flux: 
$f_{0.3-10} = (2.0 \pm 1.0) \times 10^{-15}$ erg s$^{-1}$ cm$^{-2}$ 
on 2000 June 20 ($t = 2271$ d after the explosion), 
$f_{0.3-10} = (1.7 \pm 0.7) \times 10^{-15}$ erg s$^{-1}$ cm$^{-2}$ 
on 2001 June 23 ($t = 2639$ d), and
$f_{0.3-10} = (0.9 \pm 0.4) \times 10^{-15}$ erg s$^{-1}$ cm$^{-2}$ 
on 2003 August 7 ($t = 3416$ d).
The spectrum is soft 
($f_{0.3-2} = (0.92 \pm 0.02) \times f_{0.3-10}$), and well fitted 
with optically-thin thermal plasma emission at $kT = (0.9 \pm 0.1)$ keV.
The corresponding unabsorbed luminosity on 2000 June 20 was 
$f_{0.3-10} = (1.8 \pm 0.9) \times 10^{37}$ erg s$^{-1}$.
Subsequent observations were either too short for reliable measurements 
(ObsIDs 12562 and 12568), or the source had become too faint to be 
significantly detected in individual observations. 
However, after stacking the seven {\it Chandra} observations 
from 2012 September--October ($t \approx 6750$ d), 
we detected it at an average observed flux 
$f_{0.3-10} = (4 \pm 1) \times 10^{-16}$ erg s$^{-1}$ cm$^{-2}$, 
corresponding to an emitted luminosity $\approx 4 \times 10^{36}$ erg s$^{-1}$.

The X-ray luminosity evolution at late times 
is a function of the progenitor wind 
properties as well as the SN shell expansion velocity 
\citep{FLC96,Immler+02}:
\begin{equation}
L_{\rm X} = \frac{4}{\pi (\mu m_{\rm p})^2}\,\Lambda(T)\,
           \left(\frac{\dot{M}}{v_{\rm w}}\right)^2 \, 
           \left(v_{\rm s}t\right)^{-1},
\end{equation}
where $(\mu m_{\rm p})$ is the mean mass per particle 
($\approx2.1 \times 10^{-24}$ g for a H+He plasma), 
$\Lambda (T)$ is the cooling function ($\approx3 \times 10^{-23}$ erg 
cm$^{3}$ s$^{-1}$ at 1 keV),
$\dot{M}$ is the mass loss rate in the stellar wind in the last few 
$10^4$ years before the explosion,
$v_{\rm w}$ is the stellar wind speed ($\sim10$ km s$^{-1}$), 
$v_{\rm s}$ is the shell expansion speed (for SN 1994I, 
$v_{\rm s} \approx 16,500$ km s$^{-1}$: Filippenko et al. 1995), 
and $t$ the time after the explosion.
In the simplest scenario, for constant $\Lambda (T)$, $(\dot{M}/v_{\rm w})$ 
and $v_{\rm s}$, the X-ray luminosity evolves as $t^{-1}$.
Using three {\it ROSAT} observations, and the first two {\it Chandra} 
observations, \cite[Fig. 2]{Immler+02} found an X-ray flux decline 
$f_{\mathrm{x}} \propto t^{-s}$ with $1 \la s \la 1.5$, suggesting 
a mass-loss rate $\dot{M} \sim 10^{-5}\,M_{\odot}\,\mathrm{yr}^{-1}$.
We can now better constrain the late-time decline rate, by adding 
the flux datapoints from the later {\it Chandra} observations.
We find that $f_{\mathrm{x}} \propto t^{-1.5}$, which is similar 
to the power-law decrease of the radio flux density.

\subsubsection{SN~2011dh}

\textbf{Radio.} Our VLBI observations were conducted 159 days after the explosion date (2011 May 31, hereafter $t_{0}$) of SN~2011dh. The observations detect a compact unresolved source with a flux density of 4.306 $\pm$ 0.024 mJy at the position of SN~2011dh (source J133005+471010). 

The position of our peak flux density is offset from the VLBI positions reported by \citet{Vidaletal2011} at 22~GHz 14 days after $t_{0}$ and \citet{Bietenholzetal2012} at 8.4~GHz 179 days after $t_{0}$, by 1.9 mas and 2.3 mas, respectively. The positional measurements were obtained using the same phase reference source, with the same referenced position. The offset in position with the higher frequency VLBI is less than half the synthesised beam of our 18~cm EVN observation.

With EVN observations at 8.4~GHz 179 days after $t_{0}$, \citet{Bietenholzetal2012} estimated the radius of the shock front of SN~2011dh to be 0.25$\pm$0.08 mas. The radius of the shock front was also estimated up to 92 days after $t_{0}$ via the radio spectral energy distribution, fitted to a synchrotron self-absorption (SSA) model \citep{Krauss2012, Bietenholzetal2012}. Combining the results of both observations, the radius of SN~2011dh, $r_{\mathrm{2011dh}}$, is found to follow a power-law, given by $r_{\mathrm{2011dh}}=5.85\times 10^{15} (t_{age}/30 \mathrm{days})^{0.92}$ cm. At $t_{age}$ = 159 days, $r_{\mathrm{2011dh}}$ = 2.71$\times 10^{15}$~cm $\approx$ 0.11~mas, which is $<0.1\%$ of the synthesised beam of our EVN observation.

\begin{figure}
\centering
\includegraphics[scale=0.38]{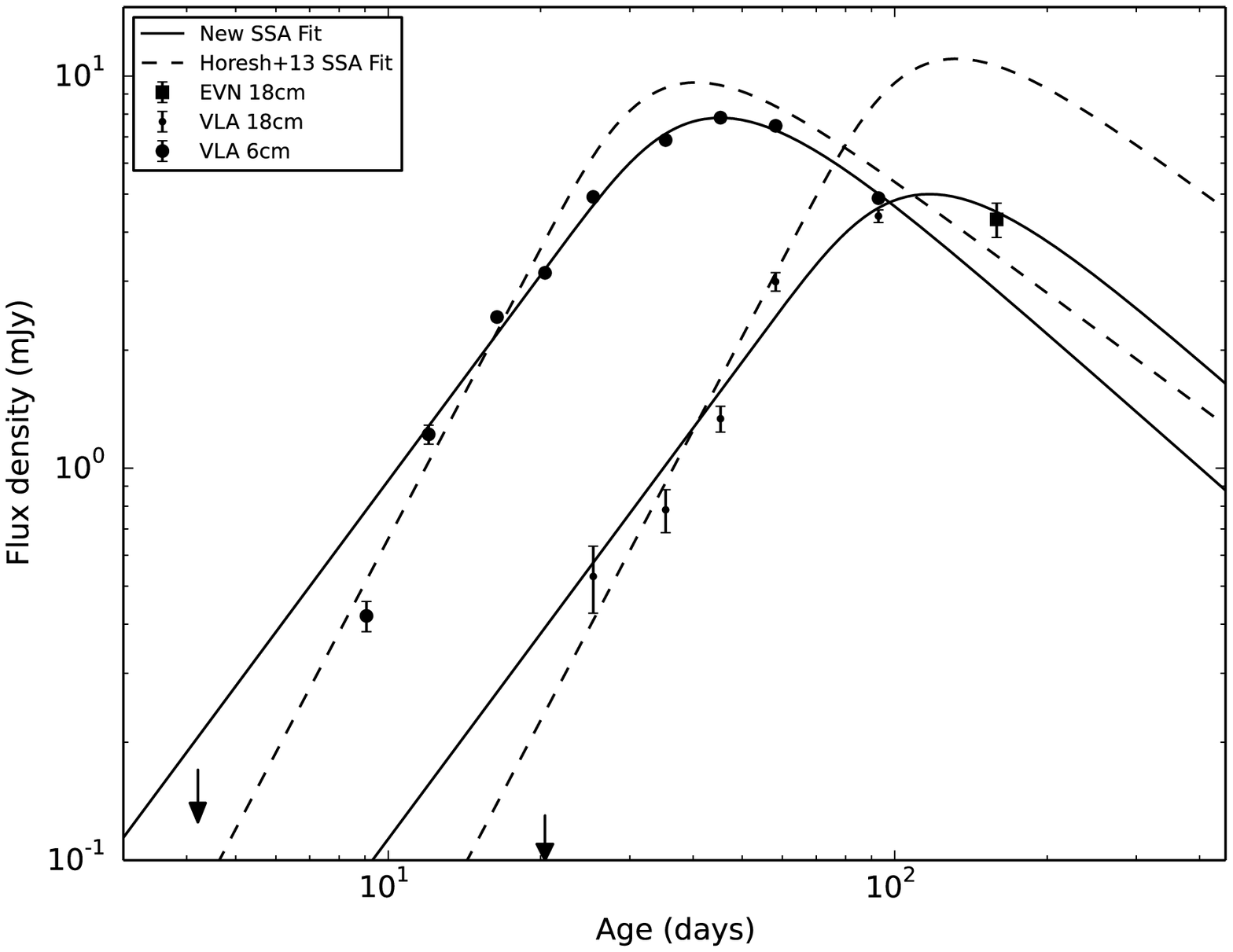} 
\caption{Light curve of SN~2011dh at 5~GHz (large filled circles) and 1.65~GHz (small filled circles and square). The 5~GHz flux density was obtained from the JVLA \citep{Krauss2012,Horeshetal13}, while the 1.65~GHz flux density data are from the JVLA (small filled circles) and the EVN observation (square) from this study. The lines are least-square fits to \Eqn~\ref{eq:weiler}, from this study (solid lines) and from \citet[dashed lines]{Horeshetal13}. The arrows are upper limits on the flux density (left - 5 GHz and right - 1.65 GHz)from \citet{Horeshetal13} obtained with the JVLA at these times.}
\label{fig:SNcurve}
\end{figure}

SN~2011dh has been observed intensively with the JVLA across several frequency bands between 1.4 to 43 GHz \citep{Krauss2012,Horeshetal13} between $t_{age}\,=\,4$ to $t_{age}\,=\,93$ days after $t_{0}$. With a detection of SN~2011dh at 159 days after $t_{0}$, it may be possible to provide constraints to the light curve model of \citet{Horeshetal13}. \Fig~\ref{fig:SNcurve} includes the 4.9/5~GHz data from \citet{Krauss2012} and \citet{Horeshetal13}, with the 1.65~GHz flux density values determined via the spectral index between the 1.4 and 1.8~GHz  flux densities at each epoch from \citet{Krauss2012}. The lines are model fits to the light curves, which are described below.

The radio emission from a supernova (SN) arises through the interaction of the SN ejecta with the circumstellar medium (CSM) \citep{Weileretal86, Weileretal02}. The resulting radio light curve is then a result of the competing effects from the declining non-thermal radio emission and the more rapid declining thermal (free-free) and non-thermal synchrotron self-absorption  (SSA) as the ejecta propagates through the CSM. This relationship is generalised in a parametric model by \citet{Weileretal02}, which allows for power-law variations in key quantities and can be written in the simplified form,

\begin{equation}
F \,=K1\,\left(\dfrac{\nu}{5\,\mathrm{GHz}}\right) ^{\alpha}\,\left(\dfrac{t_{age}-t_{0}}{1\,\mathrm{day}}\right)^{\beta}\,\left(\dfrac{1-e^{\tau_{\mathrm{opt}}}}{\tau_{\mathrm{opt}}}\right) \mathrm{mJy},
\label{eq:weiler}
\end{equation}

where $\alpha$ describes the frequency dependency (or spectral index),  $K1$ is a scaling factor and is related to the flux density at 5~GHz 1 day after the explosion, and  $\beta$ describes the power-law dependency of the flux density with the SN age, $t_{\mathrm{age}}$ \citep{Weileretal02}. The absorption term, $\tau_{\mathrm{opt}}$, describes an internal absorption by material mixed with the emitting component, which assumes a planar geometry \citep{Weileretal02,Horeshetal13}. Both the early light curve ($t<15$ days) and the radio spectra ($t<93$ days) were found to be consistent with absorption due to SSA \citep{Krauss2012,Horeshetal13}.  The absorption mechanism in \Eqn~\ref{eq:weiler} is therefore taken to be of the non-thermal SSA form and is defined by the optical depth, $\tau_{\mathrm{opt}}$, given as,

\begin{equation}
\tau_{\mathrm{opt}}\,=K5\,\left(\dfrac{\nu}{5\,\mathrm{GHz}}\right)^{\alpha-2.5}\,\left(\dfrac{t_{\mathrm{age}}-t_{0}}{1\,\mathrm{day}}\right)^{\delta},
\label{eq:SSAtau}
\end{equation}

where $K5$ is the optical depth 1 day after the explosion and $\delta$ describes the power-law dependency of $\tau_{\mathrm{opt}}$ with the SN age, $t_{\mathrm{age}}-t_{0}$ \citep{Weileretal02}. 

\medskip

The dashed lines in Fig~\ref{fig:SNcurve} are the fit to \Eqn~\ref{eq:weiler} derived by \citet{Horeshetal13}, using 22.5~GHz JVLA data for $t<15$ days. While the model from \citet{Horeshetal13} provides a better fit to the observed data for the early rise of the light cure, it clearly over-estimates the flux density for the later times. Using the 4.9/5~GHz data, a new SSA model fit to the light curve was obtained for \Eqn~\ref{eq:weiler}, which provides a better overall fit to the data at 1.65~GHz, including the turnover observed with the EVN. The final fit parameters are $\alpha$ = -1.30, $\beta$ = -1.14, $\delta$=-2.88,  $K1$ = 928.5 mJy, and $K5$= 5.5$\times\,10^{4}$ and are displayed in \Fig~\ref{fig:SNcurve} (the solid lines).

\begin{figure}
\centering
\includegraphics[scale=0.38]{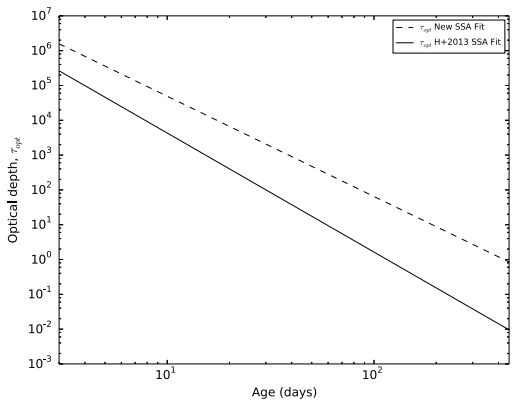}
\caption[SSA optical depth comparison between \citet{Horeshetal13} and this study]{SSA optical depth, $\tau_{opt}$, obtained via \Eqn~\ref{eq:SSAtau} using the parameters by \cite{Horeshetal13}, (solid line)  and this study (dashed line) at 1.65~GHz.}
\label{fig:SNoptdepth}
\end{figure}

The difference between the model fits can be explained through the estimates of $\tau_{\mathrm{opt}}$, which are displayed in \Fig~\ref{fig:SNoptdepth}.  At the early stages of the SN expansion, the CSM is optically thick ($\tau_{\mathrm{opt}}\gg 1$) and gradually becomes optically thin ($\tau_{\mathrm{opt}}\leq 1$) as the SN expands into the surrounding ISM \citep{Weileretal02}, coinciding with the emission of the peak radio brightness. Furthermore, the time at which the medium becomes optically thin is dependent on the frequency of emission. Thus, in deriving the model fit using the 22.5 GHz data, \citet{Horeshetal13} underestimated the optical depth (\Fig~\ref{fig:SNoptdepth}, blue dashed lines) and, therefore, the attenuation of the flux density at the lower frequencies.

 \textbf{X-Rays.} SN~2011dh was clearly detected for all observations following the explosion date. Similar to SN~1994I, the flux of SN~2011dh was found to be decreasing with time. The \textit{Chandra} data listed in \tab~\ref{tab:Chnadra_data} were used extensively by \cite{Maeda+14} to investigate the long-lasting X-ray emission from SN~2011dh and the mass loss history of the yellow supergiant progenitor. We find that our flux measurements are consistent with their reported values.

\subsection{Background Radio Sources}

\subsubsection{J133011+471041}

The radio source J133011+471041 was detected with the VLA at both 20 cm (500$\pm$22 \mujybm) and 6 cm (482$\pm$50 \mujybm) by \citet{Maddoxetal07}, giving a flat radio spectrum ($\alpha$ = -0.03). Comparing the radio position with archived \textit{Chandra} X-ray data, they found a 10$^{37}$ ergs~s$^{-1}$ counterpart, making it the brightest X-ray/radio overlap in their survey. However, no H$\alpha$ emission was found at the source position from the \textit{HST} ACS archive, leading \citet{Maddoxetal07} to suggest that J133011+471041 may be a microquasar in a radio-loud state during the time of the VLA observations. If indeed this source is a microquasar, its luminosity at 5~GHz would be a factor of two more luminous than the most luminous discovered to date (GRO J1655-40: \citealt{HR95}, \citealt{Tingayetal1995}).

\begin{figure}
\centering
\includegraphics[trim=0cm 1.7cm 0cm 0cm,clip=false,angle=-90,scale=0.35]{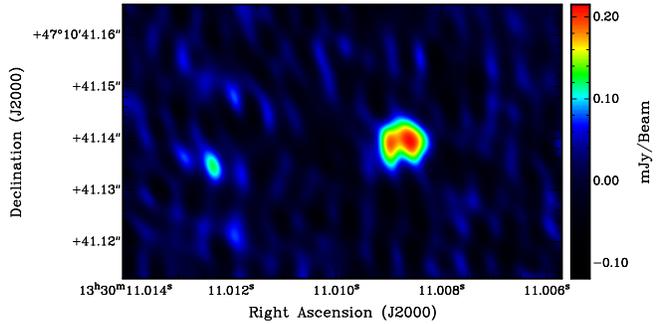} 
\caption[Higher resolution EVN image of 3011+1041]{A higher resolution EVN image of 3011+1041 made with Briggs robustness = 0, cellsize of 0.1 mas. Beam size = 6.23~mas $\times$ 3.19 mas, position angle = 13.7\deg and image RMS = 21.9 \mujybm.}
\label{fig:mgHighres}
\end{figure}

The VLBI image (\Fig~\ref{fig:VLBI}d) reveals a compact, relatively unresolved source with a total EVN flux density (see \Tab~\ref{tab:VLBI_Sources}) matching the VLA flux density within measurement errors. Since there is little or no extended emission resolved by our VLBI observations, the radio emission seen with the VLA and EVN is emanating from the same compact region, a few mas in size. Re-imaging the VLBI data with a higher resolution by setting the Briggs robust parameter to 0 and adopting a cellsize of 0.1~mas, reveals a double morphology, resembling a core-jet structure typically seen in radio galaxies or AGNs. The resulting image is shown in \Fig~\ref{fig:mgHighres}, with the peak surface brightness of the components of 270 \mujybm  ~and 233 \mujybm, separated by 4.3 mas. To the east of the double, at a distance of 38.6 mas from the brightest component of 3011+1041, is a weak unresolved source with peak surface brightness of 123~\mujybm ~that may be a jet hotspot related to J133011+471041.

\begin{figure}
\centering
\includegraphics[angle=0,scale=0.3]{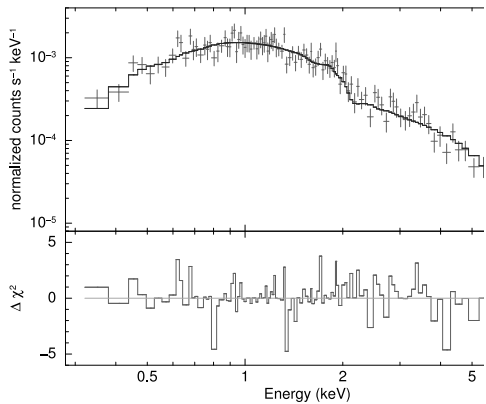} 
\caption{Stacked {\it Chandra}/ACIS-S X-ray spectrum of J133011+471041 
fitted with a power-law model ($\Gamma = 1.7 \pm 0.1$, 
and line-of-sight absorption), 
and corresponding $\chi^2$ residuals ($\chi^2_{\nu} = 100.23/105 = 0.95$).
}
\label{fig:xray_spec}
\end{figure}

\textbf{X-ray Spectrum.}  The stacked {\it Chandra} spectrum (855 ks) is well fitted with a simple 
power-law (\Fig~\ref{fig:xray_spec}) with photon index\footnote{$\Gamma$ 
is related to the spectral index $\alpha$ commonly used in radio astronomy 
by $\Gamma =  1+ \alpha$, where $S\propto\nu^{-\alpha}$.}  
$\Gamma = 1.68^{+0.10}_{-0.06}$, and line-of-sight Galactic absorption 
$N_{H} = 2 \times 10^{20}$~cm$^{-2}$. Thermal models such as disk black-body 
or thermal plasma are ruled out; adding a soft thermal component 
to the power-law model does not improve the fit.
The observed flux of the stacked spectrum 
is $f_{0.3-10} = (2.4 \pm 0.1) \times 10^{-14}$ erg s$^{-1}$ cm$^{-2}$, 
from which we infer an emitted luminosity 
$L_{0.3-10} = (2.1 \pm 0.1) \times 10^{38}$ erg s$^{-1}$ 
if the source belongs to M51a. In fact, we suggest that this source 
is more likely to be a background AGN.

We can estimate the likelihood that J133011+471041 is a background source 
using the fundamental-plane relation \citep{Merlonietal03,Plotkinetal12}.
If it were in M51a, $\log L_\mathrm{R}(5{\mathrm{GHz}}) \approx 34.2$ 
and $\log L_{2-10} \approx 38.1$. Using the best-fitting coefficients 
from \cite{Plotkinetal12}, the resulting black hole mass would be 
$\approx10^{6}\,\mathrm{M}_{\odot}$, inconsistent with an X-ray binary or even 
an intermediate-mass black hole in M51a.
The only possibility for J133011+471041 to be located in M51a is 
if its radio flux is strongly Doppler boosted, which we cannot yet rule out.

\medskip

\textbf{Optical/X-Ray.} The ratio between X-ray and optical luminosity is considered to be a reliable indicator of X-ray source classification \citep{Comastri04}. The relationship is defined as (see \citealt{Maccacaroetal1988,Bargeretal02,McHardyetal03}):

\begin{equation}
\mathrm{log}_{10}(F_{\mathrm{X}}/F_{\mathrm{opt}}) = \mathrm{log}_{10}F_{\mathrm{X}} + 5.5 + R_{\mathrm{mag}}/2.5,
\label{eq:xray/opt}
\end{equation}

where $F_{\mathrm{opt}}$ is the optical flux (in cgs) and $R_{\mathrm{mag}}$ is the Johnson-Cousins R-band flux density. The distribution of $\mathrm{log}_{10}(F_{\mathrm{X}}/F_{\mathrm{opt}})$ for spectroscopically identified X-ray AGN from 
\textit{ROSAT} \citep{Hasingeretal98}, \textit{ASCA} \citep{Akiyamaetal03}, \textit{Chandra} \citep{Giacconietal01}, and \textit{XMM-Newton} \citep{Mainierietal02,Fioreetal03} surveys fall within -1$<\mathrm{log}_{10}(F_{\mathrm{X}}/F_{\mathrm{opt}})<1$, while for stellar mass sources, such as x-ray-binaries and microquasars, $\mathrm{log}_{10}(F_{\mathrm{X}}/F_{\mathrm{opt}})>2$. 

Since the galaxy was not observed in the R-band with the HST/ACS, to estimate the $R_{\mathrm{mag}}$ of J133011+471041, the fluxes of the HST B, V and I bands were first estimated as described in \Sect~\ref{sec:AncillaryData}. The R-band flux (10.26 $\mu$Jy) was then estimated via a non-linear least squares fit to the flux densities of the B, V and I bands, from which a value of 21.1 for R$_{\mathrm{mag}}$ was determined. Substituting the values into \eqn~\ref{eq:xray/opt} results in $\mathrm{log}_{10}(F_{\mathrm{X}}/F_{\mathrm{opt}})$ = -0.03, indicating that 3011+1041 is likely a background AGN and is consistent with what was found using the fundamental-plane argument and the X-ray spectrum.


\textbf{Radio Loudness and AGN Classification.} Since we have concluded that 
J133011$+$471041 is a background source, based upon its optical/radio/X-ray flux 
ratios, we can now classify its AGN type from its radio loudness parameter 
$R_{L}$, defined as \citep {TW03, Ho2008}:
\begin{equation}
R_{L} =\dfrac{\nu F_{\nu}(5~\mathrm{GHz})}{f_{X}(2-10~\mathrm{keV})}.
\label{eq:RX}
\end{equation}
Substituting $F_{\nu}(5~\mathrm{GHz})$ = 482 $\mu$Jy 
and $f_{2-10} = 1.5 \times 10^{-14}$ erg s$^{-1}$ cm$^{-2}$ gives 
$\log R_{L} \approx -2.8$. Therefore, J133011+471041 falls within the parameter 
space of radio-loud low-luminosity AGNs (see \Fig~4 of \citealt {TW03} 
and \Fig~10 of \citealt{Ho2008}).

\subsubsection{J133005+471035}

The radio source J133005+471035 is resolved into four components with the EVN at 18~cm (\Fig~\ref{fig:VLBI}b), where the components are labelled C1 to C4 in order of decreasing peak surface brightness.


\begin{figure*}
\centering
\includegraphics[scale=0.9]{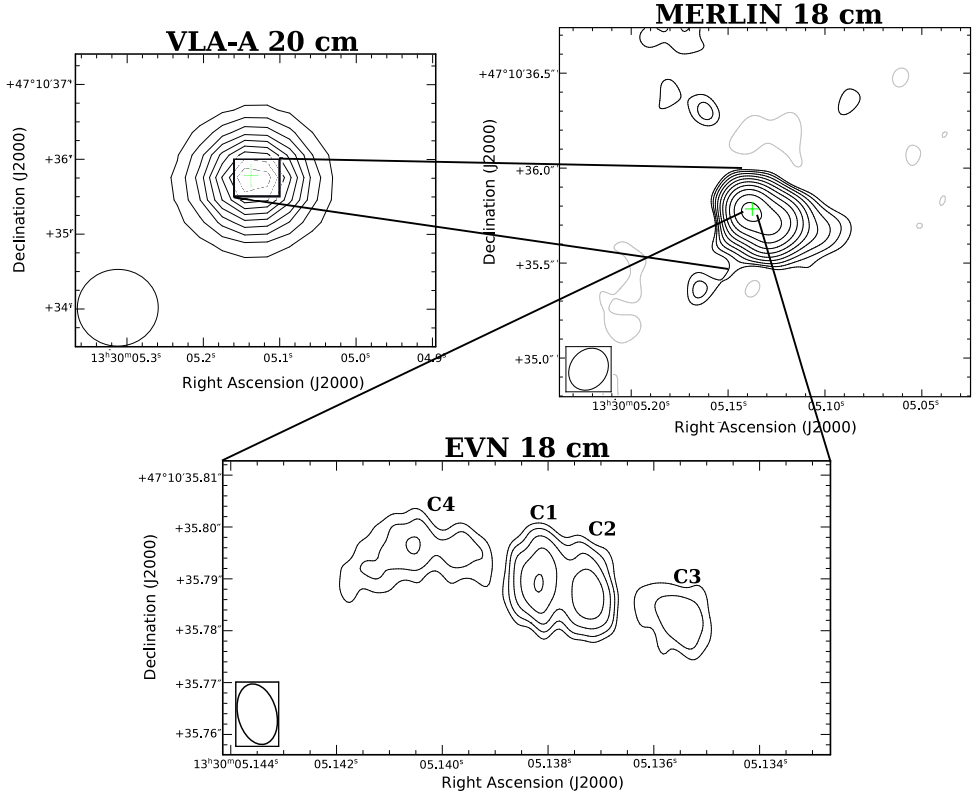} 
\caption{Radio images of the radio source J133005+471035.  \textit{VLA-A}: Contours begin at $\pm10\%$ the peak flux density (11.8 \mjybm), and increases in increments of $10\%$ of the peak flux density. \textit{MERLIN}: Image was made with the full MERLIN array (i.e. including the long baselines to Cambridge), which resulted in a beam size of 0.196\arcs $\times$ 0.164\arcs. The contours begin at $\pm3\sigma_{\mathrm{image}}$ and increase with multiples of $\sqrt{2}$, where $\sigma_{\mathrm{image}}$ = 60.3 \mujybm.  \textit{EVN}: The image is a zoomed in version of the one shown in \Fig~\ref{fig:VLBI}b, where C1 - C4 are the different components, labelled in order of decreasing peak surface brightness.}
\label{fig:kgmontage}
\end{figure*}

\begin{center}
\begin{table*}\footnotesize
\begin{minipage}{\textwidth}
\caption{VLBI detected components of J133005+471035. \emph{Columns:} (1) VLBI component defined in \Fig~\ref{fig:kgmontage}; (2) $\&$ (3) VLBI source position; (4) Angular separation of the $F_{P}$ of each component from the $F_{P}$ of C1; (5) Peak flux density; (6) Integrated flux density; (7) Size of the major axis from \textsc{imfit}; (8) Brightness temperature at $\nu$ MHz derived  from the Rayleigh-Jeans formalism: $(c^{2}F_{I})/(2\pi B_{\mathrm{maj}}k\nu^{2}$), where $c$ is the speed of light  and $k$ the Boltzmann constant.}
\label{tab:C3005}
\hfill{}
\begin{tabular}{cccccccc}
\hline\hline
\noalign{\smallskip}
 &  \multicolumn{2}{c}{Position of F$_{\mathrm{P}}$}  &  &  & &  &  \\ 
 \cline{2-3}
VLBI  &  RA & Dec  & $\Theta_{\mathrm{C1}}$ &  $F_{P}$ & $F_{\mathrm{I}}$ & B$_{\mathrm{maj}}$ & log$_{10}$ T$_{\mathrm{B}}$ \\ 
 Source &(13$^{\mathrm{h}} $30$^{\mathrm{m}}$) & (47\deg 10\arcm)& (mas) & (\mujybm) &  ($\mu$Jy) & (mas) & (K) \\ 
 (1) & (2) & (3) & (4) & (5) & (6) & (7) & (8) \\ 
  \hline
C1 & 5.1381(2) & 35.7896(4) & 0 & 211.7$\pm$25.6 & 457.2$\pm$65.2 & 12.8 & 6.26 \\ 
C2 & 5.1373(3) & 35.787308 & 9.72 & 199.8$\pm$24.6 & 414.0$\pm$59.1 & 10.98 & 6.34 \\ 
C3 & 5.1355557 & 35.782491 & 27.5 & 99.1$\pm$17.3 & 313.0$\pm$65.3 & 14.65 & 5.98 \\ 
C4 & 5.1404641 & 35.794635 & 24.2 & 91.8$\pm$16.6 & 650.2$\pm$128.6 & 32.03 & 5.61 \\ 
\hline
\end{tabular}
\hfill{}
 \end{minipage}%
\end{table*}
\end{center}
\vspace{-2.em}

\Tab~\ref{tab:C3005} lists the different properties of the  four components, including the position of the peak surface brightness, peak and total flux densities, size of the component's major axis and brightness temperature. The VLBI morphology of J133005+471035 shows evidence for a two-sided jet emanating from an active nucleus. However, without higher frequency VLBI observations it is difficult to determine which is the core that hosts the active nucleus. The similarity of both C1 and C2, in terms of size, flux density and brightness temperature suggests that they are, most probably, radio lobes/hotspots resulting from jets emanating from an unresolved, fainter active nucleus. 

In lower resolution radio maps of M51a with the VLA and MERLIN, J133005+471035 is the brightest radio source in the M51 system. \Fig~\ref{fig:kgmontage} shows the VLA-A 20~cm and MERLIN 18~cm maps in relation to the EVN 18~cm map (represented by the green crosses). In the FIRST survey the source is unresolved at 20~cm with a total flux density of 12.64~$\pm$~0.14 mJy, while at higher resolution VLA-A,  a total flux density of 13.38~$\pm$~1.50 mJy is recovered from the unresolved 20~cm VLA-A image in \Fig~\ref{fig:kgmontage}. This is similar to the peak flux density obtained by \citet{Maddoxetal07}: 9.6 \mjybm ~at 20~cm and 4.3 \mjybm ~at 6~cm, giving an intermediate spectral index of $-$0.66. The MERLIN 18~cm image displayed in \Fig~\ref{fig:kgmontage} shows an elongated structure, closely resembling the one-sided core-jet morphology of AGNs and quasars on kpc-scales \citep{BP84}. The total flux density as measured (via \textsc{blobcat}) from the MERLIN image is 9.95~$\pm$~1.00 mJy, recovering $\sim80\%$ of the flux density of the VLA-A at 18~cm (after correcting the 20 cm flux density with a spectral index of -0.66). Subtracting the total VLBI flux density listed in \Tab~\ref{tab:C3005} from the MERLIN 18~cm flux density gives 8.12~$\pm$ 1.01 mJy, which may result from thermal and non-thermal emission from star formation and non-thermal emission from kpc-scale jets.

\begin{figure*}
\centering
\includegraphics[scale=0.8]{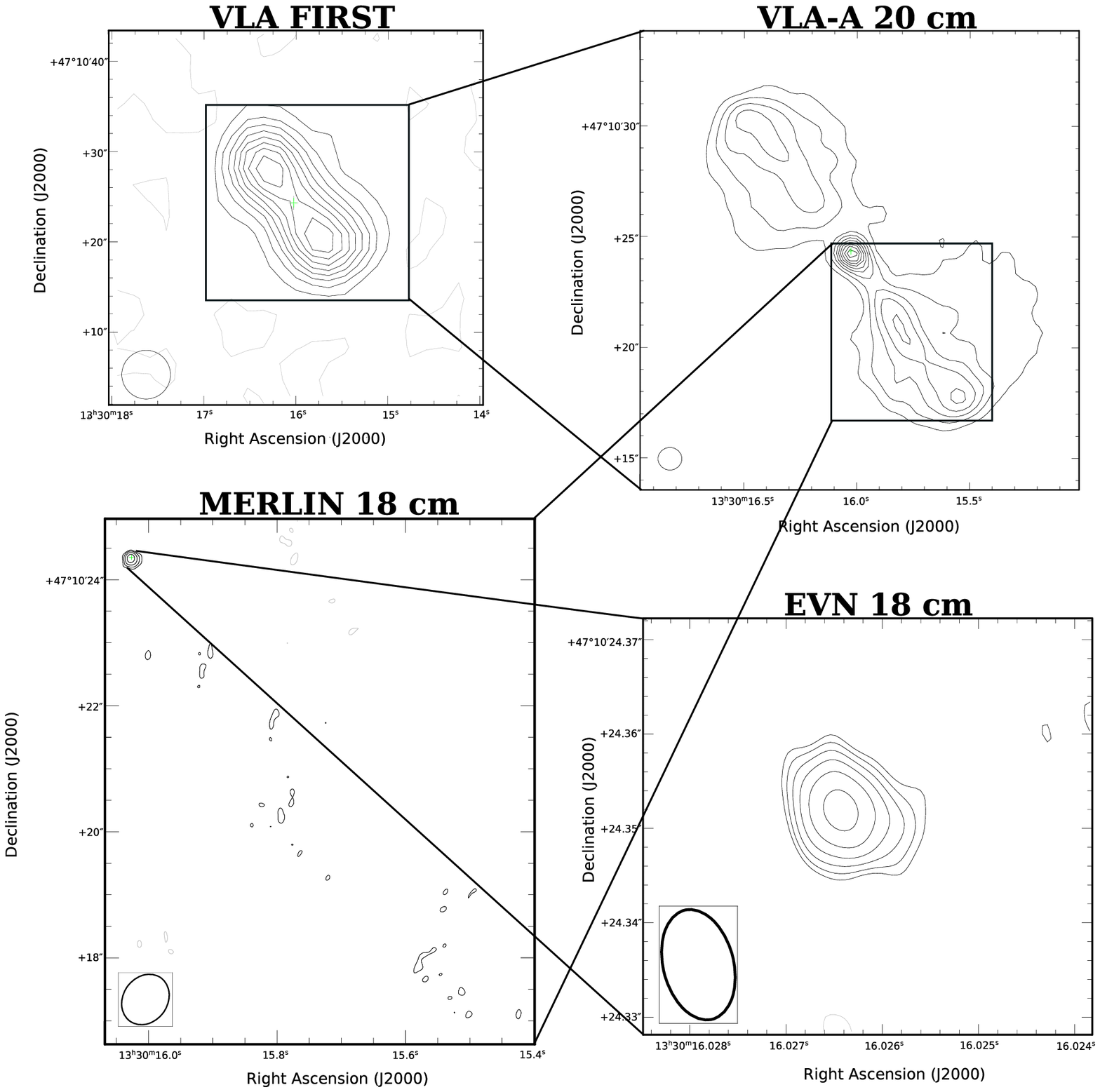} 
\caption{Radio images of the source J133016+471024. \textit{VLA}: the contours of both the FIRST and VLA-A images begin at  $\pm10\%$ the peak flux density (6.02 \mjybm $\&$ 0.982 \mjybm, respectively), and increases in increments of $10\%$ of the peak flux density.  \textit{MERLIN}: the first contours are at $\pm3\sigma$, and increases in multiples of $\sqrt{2}$, where $\sigma$ = 50$\mu$Jy beam$^{-1}$; \textit{EVN}: a zoomed in version of the one shown in \fig~\ref{fig:VLBI}.}
\label{fig:F2montage}
\end{figure*}

A faint X-ray source was found coincident with the position of J133005+471035 
in the stacked {\it Chandra} image (855 ks). 
We measure a 0.3--10 keV count rate of 
$6.0 \pm 1.2 \times 10^{-5}$ ct s$^{-1}$, mostly distributed around $\sim$1 keV.
Using Cash statistics, we obtain that the spectrum is best fitted 
by optically-thin thermal plasma emission, 
at $kT = 0.7_{-0.4}^{+0.3}$ keV, absorbed by a column density 
$N_{\rm H} \approx 5 \times 10^{21}$ cm$^{-2}$ (probably due to the location 
of the source in a dust-rich spiral arm). A power-law model 
is ruled out because it would require an unphysically steep 
photon index $\Gamma > 5$.
The observed flux corresponding to the best-fitting thermal-plasma model is 
$f_{0.3-10} = 2.1^{+0.8}_{-0.6} \times 10^{-16}$ erg s$^{-1}$ cm$^{-2}$, 
and the unabsorbed luminosity is 
$L_{0.3-10} = 0.8^{+1.2}_{-0.6} \times 10^{37}$ erg s$^{-1}$ 
if located in M51a.
The thermal nature of the source makes it consistent either 
with an SNR, if the source is in M51a, or with knots of hot gas 
shock-ionized by a jet.

\subsubsection{J133016+471024}

VLA-A 20~cm observations of the radio source J133016+471024 shows a morphology (see \Fig~\ref{fig:F2montage}) similar to nearby powerful FR~II radio galaxies such as Cygnus~A \citep{CB96} and Pictor~A \citep{Perleyetal97}, displaying the morphological components common to this type of galaxies. These sources are composed of a compact, high surface brightness temperature, typically flat spectrum source at the centre, from which elongated radio jets end in high surface brightness hotspots \citep{CB96}. The source was detected as a partially resolved double radio source in the FIRST survey with total flux densities of 12.56$\pm$0.63 mJy and 11.67$\pm$0.58 mJy for the southern and northern lobes, respectively. The higher resolution VLA-A 20 cm image resolves the double sources into two radio lobes and a central radio source, with evidence for elongated radio jets ending in hotspots at the extremities of the radio lobes. The VLA-A image recovered 82$\%$ and 71$\%$ of the FIRST flux density of the southern and northern lobes, respectively. The central source was found to have a total flux density of 2.15$\pm$0.04 mJy at 1.4~GHz\footnote{The 1.4~GHz dataset was re-imaged with the same beamsize as the 5~GHz dataset.} and 0.98$\pm$0.01 mJy at 5~GHz, giving a spectral index of -0.62.

The central radio source is unresolved with the higher resolution MERLIN 18~cm observations, with total flux density of  0.97$\pm$0.14 mJy. However, there is evidence for low surface brightness emission ($\sim$150 \mujybm) along the direction of the southern lobe/jet (see \Fig~\ref{fig:F2montage}), and at the location of the peak VLA-20 cm surface brightness component of the northern lobe (not shown in \Fig~\ref{fig:F2montage}). The VLBI observation only detects the central radio source, showing a slightly elongated structure extending from the compact source in the direction of the southern jet, suggesting that the southern jet/lobe may be relativistically beamed towards our line of sight. From the total flux density and major axis size of the central source we obtain a brightness temperature of  log$_{10}$ T$_{B}$ = 7.1, typical of AGNs (e.g. \citealt{BB98, Ulvestadetal2005}).

In the standard FR~I/FR~II classification model \citep{FR1974} both types of radio sources can be divided by radio luminosity, where the dividing luminosity is 10$^{25}$ W~Hz$^{-1}$ at 1.4~GHz (below lies the FR~I and above the FR~II galaxies). Based upon the morphology shown in \Fig~\ref{fig:F2montage}, it is possible to classify J133016+471024 as an FR~II galaxy. Assuming a minimum luminosity of 10$^{25}$ W~Hz$^{-1}$, and with a total flux density at 1.4~GHz of 2.15 mJy, J133016+471024 would be at a minimum distance of $\sim$ 6246 Mpc ($z\,\geq$ 0.95). Note, placing Cygnus A at  6246 Mpc would result in a total flux density $\sim$5 mJy at 1.4~GHz\footnote{Assuming the Flux density of Cygnus~A is $F_{1.4\mathrm{GHz}}$ = 1598~Jy at a distance of 247.2 Mpc \citep{Birzanetal04}.}.

From the Sloan Digital Sky Survey (SDSS) Data Release 9 \citep{SDSSDR9}, an optical source was found to be coincident with the position of the central radio source of J133016+471024. The R-band optical data were obtained through the SDSS Science Archive Server in the ``corrected frame" format, which have been calibrated in nanomaggies\footnote{A star of brightness 1 nanomaggie has a magnitude of 22.5 in any band, or a flux density of 3.631 \muJy. (www.sdss3.org/dr9/help/glossary.php)} per pixel, and have had a sky-subtraction applied. We performed aperture photometry with standard Heasoft packages and found a total flux density of 13.11 nanomaggies or 47.6 \muJy ~is obtained, which is equivalent to an $R_{mag}$ of 19.5.

\begin{figure}
\centering
\includegraphics[scale=0.3]{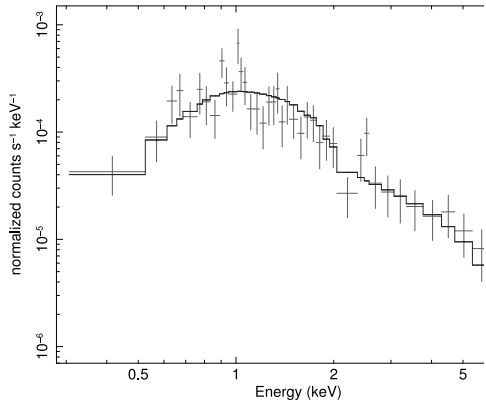} 
\caption[X-ray spectrum of the core of 3016+1024]{ 
Stacked {\it Chandra}/ACIS-S X-ray spectrum of 3016$+$1024 
fitted with a power-law model ($\Gamma = 2.0 \pm 0.2$, 
and line-of-sight absorption). The unbinned data were fitted with 
Cash statistics \citep{Cash79}, and they were then rebinned 
to a minimum signal-to-noise ratio of 2.5 for display purposes only.
}
\label{fig:F2xspec}
\end{figure}

Inspecting the stacked {\it Chandra} dataset, we found an X-ray source 
at the position of the central radio source of J133016+471024. 
We extracted a combined spectrum as described in \sect\ref{sec:xrays}.
The spectrum is best fitted (\Fig~\ref{fig:F2xspec}) with a simple power-law 
of photon index $\Gamma = 1.98^{+0.20}_{-0.19}$ and line-of-sight Galactic 
absorption $N_{\rm H} = 2 \times 10^{20}$~cm$^{-2}$; adding free intrinsic 
absorption does not improve the fit. 
The observed flux is 
$f_{0.3-10} = 4.2^{+0.6}_{-0.4} \times 10^{-15}$ erg s$^{-1}$ cm$^{-2}$.
Using \eqn~\ref{eq:RX}, we determine a radio loudness parameter 
$R_{L} = -1.97$, which places this background source firmly in the regime 
of radio-loud AGN (see \Fig~4 of \citealt {TW03} 
and \Fig~10 of \citealt{Ho2008}).

\subsubsection{J132932+471123}

While J132932+471123 is not the weakest VLBI source in our sample, 
it has the lowest S/N ratio (see \Tab~\ref{tab:VLBI_Sources}) which 
results from a high RMS noise due the source's position at the edge 
of the primary beam of Ef. Its positional offset is 3\arcm.49 in RA 
and 19\arcs.6 in Dec from the M51a centre, placing it at the edge 
of the M51a disk and beyond the survey area of \citet{Maddoxetal07}. 
At the VLBI position of J132932+471123, a radio source is detected 
in the VLA-A 20 cm image with a peak flux density of 237$\pm$49 \mujybm.
However, no corresponding source was found in the VLA-B 6~cm image 
above 60~\mujybm. 
This suggests that J132932+471123 is a steep spectrum source 
($\alpha<-1.10,\,S \propto \nu^{\alpha}$). 

In all the \textit{Chandra} observations, the source is located a few arcmin 
from the aimpoint, near the edge of the S3 chip, where the point spread 
function is larger (several arcsec) and it becomes more difficult to detect 
faint sources. Nonetheless, we find a 3$\sigma$ detection in the combined 
X-ray image (835 ks), with a 0.3--10 keV count rate of 
$(3.2 \pm 1.4) \times 10^{-5}$ ct s$^{-1}$. 
A Cash-statistic power-law fit gives a photon index 
$\Gamma = 2.0^{+1.3}_{-0.9}$ with fixed line-of-sight absorption; 
there are not enough counts for any additional fit components.
The observed flux $f_{0.3-10} = (5 \pm 2) \times 10^{-16}$ erg s$^{-1}$ cm$^{-2}$, 
which corresponds to an X-ray luminosity $\approx 5 \times 10^{36}$ 
erg s$^{-1}$ if the source is located in M51a.
No optical counterparts were found in any of the archival images used 
for this study; this is not inconsistent with an AGN identification, 
because we expect $R \approx 24.5$ mag for an X-ray-to-optical 
flux ratio of 1. 
In summary, considering its location at the outskirts of the galactic disk, 
we tentatively classify this object as a background AGN.

\subsection{Estimating the Supernova and Star Formation rates}

\subsubsection{Global supernova and star formation rates}

The supernovae rate, $\nu_{SN}$ of the entire galaxy can be estimated using the age and total number of SNRs in M51. Both estimates are given in \sect~\ref{sec:frac}, where the ages of the SNRs were estimated to be 280 - 950 yrs. As discussed in \sect~\ref{sec:frac}, our VLBI observatons detected only 3/32 radio sources in M51. Since the 3 detected radio sources are background AGNs and the AGN nucleus of M51, this suggests that the 29 undetected radio sources were SNRs that have expanded beyond the maximum size detectable with our observations. This suggests a supernovae rate range of 0.03 - 0.10 yr$^{-1}$. The global massive star-formation rate ($M\geq 5M_{\odot}$) can be estimated from the $\nu_{SN}$ using \eqn~20 from \cite{Condon92}. Thus we obtain  $SFR(M\geq 5\,M_{\odot}) \sim 0.73 - 2.43 \, M_{\odot} \mathrm{yr}^{-1}$ ($SFR(M\geq 0.1\,M_{\odot}) \sim 3.65 - 12.15 \, M_{\odot} \mathrm{yr}^{-1}$). The lower limit of our estimation is slightly higher than the SFR the total star formation rate of the entire galaxy (3.4 $M_{\odot} \mathrm{yr}^{-1}$) derived via infrared and ultraviolet observations \citep{Calzettietal05}. This suggests that some of the 29 undetected radio sources are not SNRs. A more robust estimate on $\nu_{SN}$ and the SFR from the radio observations can be achieved following the multi-epoch, multi-frequency Monte-Carlo method of \cite{Rampadarathetal2014}, requiring further observations.

\subsubsection{Star formation rate of the inner nucleus }

The star formation rate (SFR) of a galaxy is directly proportional to its radio luminosity, $L_{\nu}$ at wavelength $\nu$, assuming that all the radio emission is associated with star formation \citep{Condon92,Haarsma2000}. An estimate of the flux density associated with star formation within the inner nucleus of M51 can be obtained by subtracting the compact VLBI components from the MERLIN data. A total flux density of 2.68 $\pm$ 0.14 mJy was recovered from the MERLIN 18~cm observation, and 0.25 $\pm$ 0.02 mJy from the EVN observation. Subtraction yields a flux density  for the extended radio emission at 18~cm of 2.43 $\pm$ 0.14 mJy. However, there may be contributions to the extended emission from the jet components that were resolved-out by the EVN observations. Since this cannot be distinguished by the MERLIN observations, the flux density of the extended emission places an upper limit of the radio emission associated with star formation and hence the SFR

Using the standard relations given in \citet{Condon92}, \citet{Haarsma2000}, and \citet{Condon2002}, along with the implied standard IMFs \citep{S55,MS1979}, we derive an upper limit for the star formation rate of $SFR(M\geq 5\,M_{\odot}) < 0.044\, M_{\odot} \mathrm{yr}^{-1}$ (or $SFR(M \geq 0.1\, M_{\odot}) <0.24\, M_{\odot} \mathrm{yr}^{-1}$) for the inner nuclear region of M51, which is $\sim$10$\%$ the total star formation rate of the entire galaxy. 

%
%

\section{Conclusions}

\label{sec:conclusion}

This paper presents the deepest, widest VLBI survey of a nearby grand design spiral galaxy. The target for our study is the Whirlpool galaxy (M51a, NGC~5194), that is undergoing a merger with its smaller companion, NGC~5195. The full disk of M51 was surveyed using the multi-phase centre technique (192 phase centres) for compact radio sources. By combining this single pointing of the EVN, with multi-wavelength radio observations, X-ray and optical data we investigated: the properties of the nuclear region of M51a (including a pc-scaled jet); the light curves of the Type~IIb supernovae, SN~2011dh (radio) and the Type Ib/c supernova, SN~1994I (radio and x-ray); the nature of four background AGNs (J133005+471035, J133011+471041, J132932+471123, J133016+471024); and the supernova and star formation rates of M51a. 

It is important to note that of the 192 phase centres, only six provided positive detections, indicating that most of the computing time was spent on imaging empty fields. Future VLBI surveys of nearby galaxies would therefore benefit from limiting the phase centres to only sources detected with lower resolution interferometers (e.g. the JVLA and eMERLIN). Nonetheless, this study demonstrates the extensive scientific capabilities of a single wide-field VLBI observation.

\section*{Acknowledgements}

The International Centre for Radio Astronomy Research is a joint venture between Curtin University and the University of Western Australia, funded by the state government of Western Australia and the joint venture partners. The EVN is a joint facility of European, Russian, Chinese, South African and other radio astronomy institutes funded by their national research councils. The VLA operated by the NRAO is a facility of the National Science Foundation operated under cooperative agreement by Associated Universities, Inc. MERLIN is a National Facility operated by the University of Manchester at Jodrell Bank Observatory on behalf of STFC. The \textit{Hubble Space Telescope}, is operated by NASA and the Space Telescope Science Institute. The \textit{Chandra} X-ray telescope is operated for NASA by the Smithsonian Astrophysical Observatory. iVEC is a jointly operated venture between CSIRO, and various Western Australian universities, including Curtin University. SJT is a Western Australian Premier's Research Fellow, funded by the state government of Western Australia.  We wish to thank the Joint Institute for VLBI in Europe (JIVE) for conducting the EVN observations, Alessandra Bertarini, Helge Rottmann and Walter Alef for performing the data correlation for the EVN data at the University of Bonn, and Olaf Wucknitz for assisting with moving the EVN data from Bonn to Perth within reasonable time. Thanks to Chris Phillips for providing the support required to successfully utilise the computing cluster, CAVE. This research has made use of material from the NASA/IPAC Extragalactic Database (NED), operated by the Jet Propulsion Laboratory, California Institute of Technology, under contract with NASA, the VizieR catalogue access tool, and the Aladin Sky Atlas both operated by CDS Strasbourg, France. HR acknowledges
support from ERC-StG 307215 (LODESTONE). We wish to thank the anonymous referee for very useful comments that have improved the quality of this manuscript.

\bibliographystyle{mn2e}
\bibliography{thesisref}

\clearpage
\appendix
\section{LR Method}

The ratio used in \sect~\ref{sec:LRmethod} to estimate the likelihood  that a pixel in a VLBI image is associated with a catalogued source (the alternative hypothesis, $H_{1}$), or a noise spike (the null hypothesis, $H_{0}$) is given as,

\begin{align}
\mathrm{LR} = \dfrac{p(H_{1})}{p(H_{0})},
\end{align}
 
where p($H_{0}$) is the likelihood the pixel is a noise spike and is taken as the Cumulative Distribution Function of the positive half of the Gaussian distribution, while p($H_{1}$) is the likelihood the pixel is associated with a catalogued source and is dependent on the position of the pixel in relation to a catalogued source position/morphology as defined by \citet{Morganetal2013}. Before applying the LR method, the \textsc{aips} task \textsc{uvfix} was used to produce visibilities centred on the catalogued positions allowing the data to be averaged in frequency and time, with little loss of amplitude response due to image smearing. Dirty images were generated for each of the catalogued source positions with the full array\footnote{This test was also conducted for the reduced array of maximum baseline of 7.5~M$\lambda$. However there was no difference in the final result compared to the full array.}, using parameters: cellsize = 0.5~mas; image size listed in \Tab~\ref{tab:LRprep}; and natural weighting. In addition, the outer 512 pixels removed prior to applying the LR method. The required VLBI parameters to search for sources from both catalogues are listed in \Tab~\ref{tab:LRprep}.  

LR images were generated for each catalogued source and the position of the maximum LR value was visually compared to the corresponding VLBI peak pixel S/N. Pixels with log$_{10}\mathrm{(LR)}>10$ and $\mathrm{S/N}>6.7\sigma$ were taken as confirmed detections.  \Tabs~\ref{tab:LRresFIRST} and \ref{tab:LRresMad} list the LR values for the confirmed detections along with the corresponding VLBI image parameters for confirmed detections, and the non-detection with the highest LR value for both surveys.  As with the results of \citet{Morganetal2013}, the LR test was found to be more dependent on the VLBI pixel S/N (i.e. p($H_{0}$)), than the positional constraints given by the lower resolution catalogues. 

\begin{figure*}\footnotesize
\centering
\includegraphics[scale=0.35,trim=0cm 1cm 0cm 0cm,clip=true]{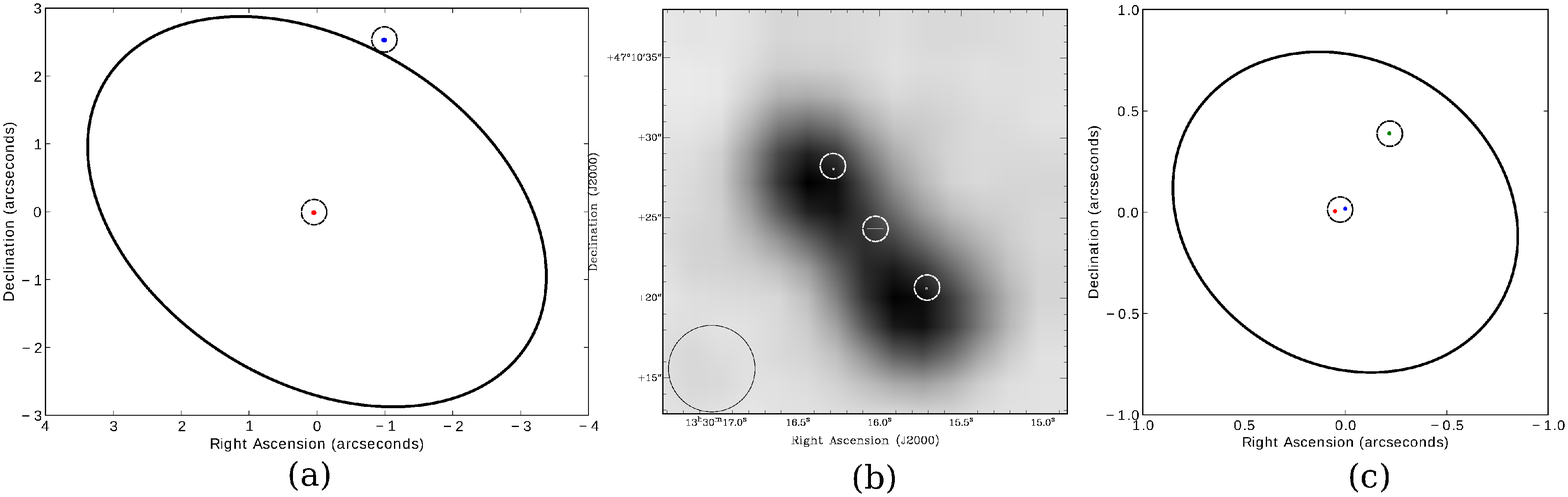}
\caption{{\scriptsize (a) The location of the peak VLBI pixel with respect to the catalogued position of the nearest corresponding FIRST source. The circles are: red (J133005+471035); and blue (J132952+471142). (b) The location of the single VLBI source (J133016+471024; white cross) located between two FIRST sources (greyscale). The ellipse gives the FIRST beam size (see \Tab~\ref{tab:LRprep}). (c) The location of the peak VLBI pixel with respect to the catalogued position of the nearest corresponding 20~cm \citet{Maddoxetal07} source. The circles are: red (J133005+471035);  green (J133011+471041); blue (J132952+471142). The ellipse in both (a) and (c) are the beam sizes of the respective surveys, centred on the catalogued position. The size of the circles indicate the astrometric error on the catalogued positions via the relationship, B$_{\mathrm{maj/min}} \cdot \sigma/2S_{P}$, where B$_{\mathrm{maj/min}}$ is the major or minor beamsize, $\sigma$ is the RMS noise in the catalogue images and $S_{P}$ is the peak flux density of the catalogued source \citep{Morganetal2013}. Dashed circles have been placed around the sources to aid the reader.}}

\label{fig:VLBIposVLA}
\end{figure*}

An example of this is displayed in \Fig~\ref{fig:VLBIposVLA}, where a $22\sigma$ VLBI pixel was discovered $\sim$6\arcs from two FIRST sources, but has log$_{10}$(LR) values $>$100 when compared with both FIRST sources. However, pixels that were found to have better positional constraints (i.e. higher values of p($H_{1}$)), nevertheless, resulted in lower LR values. As such no significant advantage is given by the LR method over simply detecting the brightest pixels and thus, it is difficult to determine whether pixels with S/N$<6.7\sigma$  are indeed real detections, and are henceforth considered as resulting from noise.

Of the three \citet{Maddoxetal07} sources detected by the EVN, two were found with very good agreement between the VLBI and lower resolution source positions (see \Fig~\ref{fig:VLBIposVLA}c). It may be possible to use estimates of the positional error of the \citet{Maddoxetal07} survey to improve the LR method and increase the chance of detecting sources below $6.7\sigma$. However, this is beyond the scope of this study.

\begin{center}
\begin{table}\footnotesize
\caption{Parameters used in the likelihood ratio method to search for sources within the EVN images. \emph{Columns:} (1) Catalogue of radio sources in M51; (2) Number of sources detected in M51 by both catalogues; (3) Major axis of the survey beam; (4)  Number of pixels per dimension required to image the field of size B$_{\mathrm{maj}}$.}
\label{tab:LRprep}
\hfill{}
{\renewcommand{\tabcolsep}{3pt}
\begin{tabular}{lccc}
\hline\hline
\noalign{\smallskip}
 & Source  & B$_{\mathrm{maj}}$ &  Image size   \\
 Catalogue& Count & ($\prime\prime$) & (pixels)   \\ 
(1) &(2) & (3) & (4) \\ 
 \hline
FIRST \dotfill & 15 & 5 & 7936  \\ 
Maddox (20~cm) \dotfill  & 96 & 1.51 & 1984   \\ 
\hline
\end{tabular}
}
\hfill{}
\end{table}
\end{center}

\begin{center}
\begin{table*}\footnotesize
\caption{Results of the likelihood ratio analysis of the VLA First catalogue. \emph{Columns:} (1) Flux density of the brightest pixel from images made with the full EVN array; (2) The 1$\sigma$ of the individual images obtained  through fitting a Gaussian distribution  to the  histogram of pixel fluxes; (3) Signal  to noise  ratio; (4) The probability the brightest pixel in the image is a noise spike, i.e. a source of  S/N ratio would occur by chance  once in every 1/$10^{-log_{10}(p(H_{o}))}$ pixels; (5) Likelihood the pixel is related to a catalogued source based upon position; (6) The likelihood ratio parameter of  the brightest pixel in the  image. See text for details.}
\label{tab:LRresFIRST}
\hfill{}
\noindent\makebox[\textwidth]{
\begin{tabular}{cccccc}
\hline\hline
\noalign{\smallskip}
 
VLBI$_{\mathrm{peak}}$ & $\sigma$ & S/N &  &    \\ 
 (\mujybm) & (\mujybm) & (VLBI$_{\mathrm{peak}}$/$\sigma$)  & $-$log$_{10}$(p(H$_{o}$)) & $-$log$_{10}$(p(H$_{1}$)) & log$_{10}$(LR) \\
  (1) & (2) & (3) & (4)  & (5) & (6)\\ 
 \hline
 326.42$^{a}$ & 11.75 & 27.77 & 169.4  & 0.49 & 168.9 \\ 
 233.00$^{b}$ & 10.03 & 23.24 &  118.9 & 0.89 & 118.0    \\ 
 231.16$^{b}$ & 10.09 & 22.90 & 115.7  &  1.09 & 114.6 \\
 84.13$^{c}$ & 8.96 & 9.39 & 20.5 & 1.09 & 19.4   \\ 
\hline
 53.35$^{\dagger}$ & 9.00 & 5.93 & 8.8 &0.60  & 8.2   \\  
\hline
\multicolumn{6}{l}{{\scriptsize Source identification (see \tab~\ref{tab:VLBI_Sources}): $^{a}$ J123005+471035; $^{b}$ A single VLBI source (J133016+471024) located between two FIRST }}\\
\multicolumn{6}{l}{{\scriptsize  sources. See \Fig~\ref{fig:VLBIposVLA}; $^{c}$ J132952+471142. $^{\dagger}$ The non-detection with the highest LR.} }\\
\multicolumn{6}{l}{ {\scriptsize (6) The likelihood ratio parameter of  the brightest pixel in the  image. See text for details.}}
\end{tabular}
}
\end{table*}
\end{center}

\begin{center}
\begin{table*}\footnotesize
\caption{Results of the likelihood ratio analysis of the Maddox catalogue. }
\label{tab:LRresMad}
\hfill{}
\begin{tabular}{cccccc}
\hline\hline
\noalign{\smallskip}
 
VLBI$_{\mathrm{peak}}$ & $\sigma$ & S/N &  &   \\ 
 (\mujybm) & (\mujybm) & (VLBI$_{\mathrm{peak}}$/$\sigma$)  & $-$log$_{10}$(p(H$_{o}$)) & $-$log$_{10}$(p(H$_{1}$)) & log$_{10}$(LR) \\ 
  (1) & (2) & (3) & (4)  & (5) & (6) \\ 
 \hline
 327.26$^{a}$ & 17.33 & 18.31 & 74.5  & 0.49 & 74.0   \\ 
 317.42$^{b}$ & 30.23 & 10.82 & 26.9 & 0.40 & 26.5\\ 
 86.07$^{c}$ & 10.12 & 8.51 & 17.0 & 0.30 & 16.7  \\ 
\hline
 47.48$^{\dagger}$ & 8.52 & 5.57 & 7.9 & 0.60 & 7.3  \\  
\hline
\multicolumn{6}{l}{{\scriptsize \textbf{Notes}. Symbols and headings have same meaning as \Tab~\ref{tab:LRresFIRST}}}\\
\multicolumn{6}{l}{{\scriptsize Source identification (see \tab~\ref{tab:VLBI_Sources}): $^{a}$ J133005+471035; $^{b}$ J133011+471041; $^{c}$ J132952+471142.}}
\end{tabular}
\hfill{}
\end{table*}
\end{center}

\end{document}